\documentclass{osa-article}

\journal{osajournal}
\articletype{Research Article}
\usepackage{lineno}
\usepackage{siunitx}
\usepackage{mathtools}
\newcommand{\AAA}{{\cal A}}
\newcommand{\BB}{{\cal B}}

\begin{document}

\title{Coherent mechanical noise cancellation and cooperativity competition in optomechanical arrays}

\author{Matthijs H.\ J.\ de Jong,\authormark{1,2,$\dagger$} Jie Li,\authormark{1,3,$\dagger$}, Claus G\"{a}rtner,\authormark{1} Richard A.\ Norte,\authormark{2} and Simon Gr\"{o}blacher\authormark{1,*}}

\address{\authormark{1}Kavli Institute of Nanoscience, Department of Quantum Nanoscience, Delft University of Technology, Lorentzweg 1, 2628CJ Delft, The Netherlands\\
\authormark{2}Department of Precision and Microsystems Engineering, Delft University of Technology, Mekelweg 2, 2628CD Delft, The Netherlands\\
\authormark{3}Zhejiang Province Key Laboratory of Quantum Technology and Device, Department of Physics, Zhejiang University, Hangzhou 310027, China\\
\authormark{$\dagger$}These authors contributed equally}

\email{\authormark{*}s.groeblacher@tudelft.nl} %% email address is required

\vspace*{0.25cm}
\homepage{This work was published in \href{https://doi.org/10.1364/OPTICA.446434}{Optica \textbf{9}, 170--176 (2022)}.} %% author's URL, if desired

\begin{abstract*} %nocopyright
Studying the interplay between multiple coupled mechanical resonators is a promising new direction in the field of optomechanics. Understanding the dynamics of the interaction can lead to rich new effects, such as enhanced coupling and multi-body physics. In particular, multi-resonator optomechanical systems allow for distinct dynamical effects due to the optical cavity coherently coupling mechanical resonators. Here, we study the mechanical response of two SiN membranes and a single optical mode, and find that the cavity induces a time delay between the local and cavity-transduced thermal noises experienced by the resonators. This results in an optomechanical phase lag that causes destructive interference, cancelling the mechanical thermal noise by up to \SI{20}{\decibel} in a controllable fashion, matching our theoretical expectation. Based on the effective coupling between membranes, we further propose, derive and measure a collective effect, cooperativity competition on mechanical dissipation, whereby the linewidth of one resonator depends on the coupling efficiency (cooperativity) of the other resonator.
\end{abstract*}

\section{Introduction}
Cavity optomechanics~\cite{Aspelmeyer2014} addresses the interaction between electromagnetic fields and mechanical motion. In recent years, multi-mode optomechanics, such as multiple mechanical resonators interacting with a common cavity field, has received significant attention and offered a platform for studying rich physics, including hybridization~\cite{Lin2010,Massel2012,Shkarin2014,OckeloenKorppi2019} and synchronization~\cite{Zhang2012, Bagheri2013,Sheng2020} of mechanical modes, mechanical state swapping~\cite{Weaver2017}, coherent~\cite{Spethmann2016} and topological~\cite{Xu2016} energy transfer, and two-mode squeezed mechanical states~\cite{Mahboob2014,Pontin2016,OckeloenKorppi2018}. In particular, optomechanical systems consisting of multiple SiN membranes have seen considerable progress towards the enhancement of their single-photon coupling rate~\cite{Xuereb2012,Li2016,Piergentili2018,Newsom2020}, and have been the subject of many theoretical proposals~\cite{Heinrich2011,Woolley2014,Li2015,Buchmann2015,Kipf2014}. Compared to the relatively simple description of the standard optomechanical system, arrays of mechanical resonators coupled to a common optical mode offer the prospect of studying complex new physical effects and the ability to achieve individual control over each constituent of a multi-element system.

In this work, we study two mechanical resonators coherently coupled to a common cavity mode, that couples the thermal mechanical noise of the two resonators in an effective mechanical beam-splitter interaction~\cite{Seok2012,Buchmann2015} that can be used to swap the mechanical states~\cite{Weaver2017,Fedoseev2021} or topologically transfer energy between them~\cite{Xu2016}. By operating in the side-band unresolved regime, the optomechanically scattered photons that mediate this effective mechanical beam-splitter interaction can remain coherent in the cavity, which adds a stochastic time delay to this process. This results in a time delay in the effective (local and transduced) noise experienced by each resonator, which causes destructive interference when the mechanical resonator spectra overlap. We measure up to \SI{20}{\deci\bel} cancellation of mechanical noise, matching well with our theoretical model. This provides a new interference mechanism distinct from that attributed to direct mechanical coupling between two resonators~\cite{Lin2010,Stassi2017}, to multiple optical modes~\cite{Dobrindt2010,Yanay2016} or optical modulation~\cite{Caniard2007}, which can clearly be excluded in our system. 

We further propose and derive another new collective effect, resulting in a cooperativity competition of the mechanical dissipation, which we also observe in our measurements. This competition arises between the dissipation dynamics of two mechanical resonators coupled to the same optical field and leads to a linewidth broadening of one resonator that depends on the optomechanical cooperativity of the other resonator.

\section{Theory and experimental setup} 
\begin{figure}[t]
\centering
	\includegraphics[width = 0.6\textwidth]{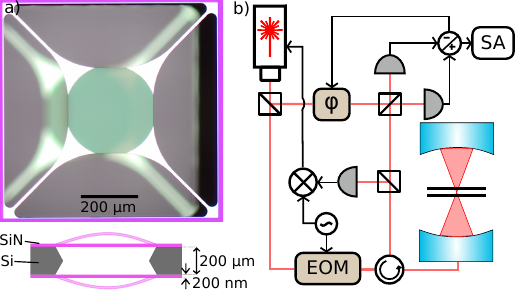}
	\caption{a) (top) Microscope image of a double-membrane device.	(bottom) Cross-section of the double-membrane chip, with the SiN membranes on either side of the Si substrate. The \SI{200}{\micro\metre} spacing between the two membranes is rigid, fixed by the substrate. b) Schematic of the experimental setup. The laser wavelength is locked to the cavity length using a Pound-Drever-Hall locking scheme and the mechanics of the membranes are measured via homodyne detection. SA: Spectrum analyzer, $\varphi$: Fiber stretcher, EOM: Electro-optic modulator.}
	\label{Figure1}
\end{figure}

Our system consists of an array of two nominally identical \SI{200}{\nano\metre} thick SiN membranes (Fig.~\ref{Figure1}(a)) with fundamental frequencies $\omega_{1,2} \simeq 2\pi \times 150$~\si{\kilo\hertz} and linewidths $\gamma_{1,2} \simeq 2\pi \times 0.1$~\si{\hertz}. They are patterned with a photonic crystal with \SI{35}{\%} reflectivity at \SI{1550}{\nano\metre}~\cite{Gaertner2018}, characterized in a previously described setup~\cite{Norte2016}. The double-membrane chip is placed close to the center of a \SI{49.6}{\milli\metre} long Fabry-Pérot cavity (free spectral range \SI{3.023}{\giga\hertz}, beam waist \SI{33}{\micro\metre}), with an empty-cavity optical linewidth (full width at half maximum) of $\kappa_\mathrm{e} = 2\pi \times 128$~\si{\kilo\hertz}, in principle putting us into the optomechanical sideband resolved regime (total linewidth $\kappa \lesssim \omega_j$). The membranes cause additional optical loss when placed inside the cavity due to scattering and small imperfections in the alignment, resulting in a linewidth $\kappa \gtrsim 2\pi \times 300$~\si{\kilo\hertz}, with a strong dependence on the exact position~\cite{Jayich2008} and alignment~\cite{Gaertner2018} of the membranes. The mechanical motion of the membranes is coupled to the optical cavity frequency $\omega_\mathrm{c}$ with vacuum optomechanical coupling rates $g_{0,1}$ and $g_{0,2}$ respectively. A laser at frequency $\omega_\ell$ is coupled to the cavity with coupling strength $E = \sqrt{P_\ell\kappa_\mathrm{e}/\hbar\omega_\ell}$, where $P_\ell$ is the laser power and $\kappa_\mathrm{e}$ the external coupling rate of the cavity.

The behavior of the membranes is investigated using a homodyne detection setup, schematically shown in Fig.~\ref{Figure1}(b), for which we lock the laser wavelength ($\lambda = 1549.62$~\si{\nano\metre}) to the cavity length using a Pound-Drever-Hall (PDH) locking scheme~\cite{Black2001}. By tuning the parameters of our PID controller, we can lock the laser beam slightly off-resonant with our cavity, and the red (blue) detuned laser can be used to cool (amplify) our optomechanical system.

The Hamiltonian describing our system is given by
\begin{equation}
\hat{H}/\hbar= \omega_\mathrm{c} \hat{a}^\dagger \hat{a} +\!\!\sum_{j = 1,2} \left(\frac{\omega_j}{2}\left( \hat{x}_j^2 \!+\! \hat{p}_j^2 \right)\!-\!g_{0,j}\hat{a}^\dagger \hat{a} \hat{x}_j \right) + iE\left( \hat{a}^\dagger e^{-i\omega_\ell t}\!\!-\!\mathrm{H.c.}\right)
\end{equation}
with $\hat{a}$ ($\hat{a}^\dagger$) the annihilation (creation) operator of the cavity mode, $\hat{x}_j$ and $\hat{p}_j$ the dimensionless position and momentum of the $j^\textrm{th}$ mechanical resonator. We are interested in the fast fluctuations of the mechanical operators ($\delta \hat{x}_j$, $\delta \hat{p}_j$) and optical field, which are described by the quantum Langevin equations (QLEs, see the supplemental document Sec.~1 for details),
\begin{equation}
\label{QLE}
\begin{aligned}
\delta \dot{\hat{x}}_j &= \omega_j \delta \hat{p}_j \\
\delta \dot{\hat{p}}_j &= -\omega_j \delta \hat{x}_j - \gamma_j \delta \hat{p}_j + G_j^* \delta \hat{a} + G_j \delta \hat{a}^\dagger + \hat{\xi}_j \\
\delta \dot{\hat{a}} &= -\left( i \Delta + \kappa/2 \right) \delta \hat{a} + i \sum_{j=1,2} G_j \delta \hat{x}_j + \sqrt{\kappa} \hat{a}^\mathrm{in}
\end{aligned}
\end{equation}
where $\hat{\xi}_j$ and $\hat{a}^\mathrm{in}$ are the mechanical and optical noise terms. We have further introduced an effective detuning $\Delta = \omega_\mathrm{c} - \omega_\ell - \sum_j \frac{g_{0,j}^2}{\omega_j} \lvert \langle \hat{a}\rangle \rvert^2$ and the effective optomechanical coupling rate $G_j = g_{0,j} \langle \hat{a} \rangle = g_{0,j} E/(\kappa/2 + i\Delta)$, with $\langle \hat{a} \rangle$ being the average cavity field amplitude. We can solve these equations by taking the Fourier transform and deriving the expected power spectral density (PSD) detected from the cavity output field (supplemental document Sec.~2).

\section{Results}

\subsection{Interference from optomechanical phase lag}
In this section, we motivate the introduction of the optomechanical phase lag, and show that it leads to interference in the dynamics of the two mechanical resonators. To simplify the following analysis, we take the phase of $\langle \hat{a} \rangle$ such that the $G_j$ are real (the full derivation keeping any complex values of $G_j$ is given in the supplemental document Sec.~1). If we solve the QLEs, Eq.~\eqref{QLE}, the position fluctuations for example for resonator 1 take the form
\begin{equation}
\label{eqx1}
\delta \hat{x}_1  =  \chi_1^{\rm eff} (\omega) \Bigg\{ \frac{ -G_1  G_2  \big[ \chi_c (\omega) - \chi_c^* (-\omega) \big] \, \chi_2 (\omega)  \, \hat{\xi}_2 +  i G_1 \! \sqrt{ \kappa} \, \big[ \hat{a}^{\rm in} \chi_c (\omega)  + \hat{a}^{\rm in, \dag} \chi_c^* (-\omega) \big] } {i +  G_2^2 \, \chi_2 (\omega)  \big[ \chi_c (\omega) -  \chi_c^* (-\omega) \big]} + \hat{\xi}_1 \Bigg\} ,
\end{equation}
where we have introduced the natural susceptibility of the mechanical resonators, $\chi_j (\omega) = \frac{\omega_j}{\omega_j^2 - \omega^2 - i \gamma_j \omega}$, and of the cavity field, $\chi_c (\omega) =  \frac{1}{ \kappa/2 + i (\Delta - \omega)}$ ($\chi^*_c (-\omega) =  \frac{1}{ \kappa/2 - i (\Delta + \omega)}$), and an effective susceptibility that incorporates the optomechanical effects on the susceptibility of the mechanical resonator,
\begin{equation}\label{chi1eff}
\chi_1^{\rm eff} (\omega) = \Bigg[ \frac{1}{\chi_1 (\omega)}  +  \frac{G_1^2 \left( \chi_c (\omega) - \chi_c^* (-\omega) \right)}{i + G_2^2 \, \chi_2 (\omega) \left( \chi_c (\omega) - \chi_c^* (-\omega) \right)}   \Bigg]^{-1}.
\end{equation}
Eq.~\eqref{eqx1} features terms that contain the different noise sources, the optical noises $\hat{a}^\mathrm{in}, \hat{a}^{\mathrm{in},\dagger}$ and the mechanical noises of both resonators, $\hat{\xi}_1, \hat{\xi}_2$. If we neglect the optical noises, which is a valid assumption if the system is at room temperature, we can see that the position fluctuations of the resonator depend on an effective mechanical noise, $\hat{\xi}^\mathrm{eff}$,
\begin{equation}
\label{Effectivenoise}
\hat{\xi}_1^\mathrm{eff}(\omega) = \hat{\xi}_1+ M_1 \hat{\xi}_2, \quad
\hat{\xi}_2^\mathrm{eff}(\omega) = \hat{\xi}_2 + M_2 \hat{\xi}_1
\end{equation}
with
\begin{equation}
\begin{aligned}
\label{Mcouplingfactors}
M_1 (\omega) &= \frac{i\chi_2(\omega) G_1 G_2 \left(\chi_c (\omega) - \chi_c^*(-\omega)\right)}
{1 - i G_2^2 \chi_2(\omega) (\chi_c(\omega) - \chi_c^*(-\omega))} \\
M_2 (\omega)&= \frac{i \chi_1(\omega) G_1 G_2\left(\chi_c (\omega) - \chi_c^*(-\omega)\right)}
{1 - i G_1^2 \chi_1(\omega) (\chi_c(\omega) - \chi_c^*(-\omega))}.
\end{aligned}
\end{equation}
This is the crucial point: The position fluctuations of any (one) of the two resonators are not only dependent on its local thermal bath, but also on the thermal bath of the other via the optical field (Fig.~\ref{Figure2}(a)), as is well-understood for general coupled resonators~\cite{Boyanovsky2017}. This cross-term between the resonators is the effective mechanical beam-splitter interaction used for state-swapping and energy transfer between the mechanical resonators\cite{Weaver2017,Xu2016,Seok2012,Buchmann2015,Fedoseev2021}. It represents photons that have been scattered with phonon transfer (i.e. optomechanically scattered) from one resonator, and subsequently re-scattered from the other. While this is a second-order optical process, it is linear in the mechanical operators, so it is not eliminated by the linearization of the QLE's (Eq.~\eqref{QLE}). We quantify the rate of this process in the supplemental document, Sec.~3, and show that the transduced noise can be similar in amplitude to the local noise.

To evaluate these expressions and obtain a power spectral density (PSD) such as the one we detect in our experimental setup, conventionally, one assumes each of the mechanical baths to be Markovian (if $Q_j = \frac{\omega_j}{\gamma_j} \gg 1$~\cite{Benguria1981,Giovannetti2001}), with autocorrelators for $\hat{\xi}_j$ as 
\begin{equation}
\label{Mechnoisecorrelation}
\left\langle \hat{\xi}_j(t)\hat{\xi}_j(t') + \hat{\xi}_j(t')\hat{\xi}_j(t) \right\rangle/2 \approx \gamma_j (2\bar{n}_j + 1)\delta(t-t'),
\end{equation}
with $\bar{n}_j$ the mean thermal phonon number (supplemental document Sec.~4). Based on Eq.~\eqref{Effectivenoise}, we can write an autocorrelator for the effective noise, which will contain terms from both thermal baths.

It is here that we introduce new physics. In Eq.~\eqref{Effectivenoise}, both noises have an \emph{immediate} effect on the position fluctuations of the resonator: $\delta \hat{x}_1 (t)$ is dependent on $\hat{\xi}_1(t)$ and $\hat{\xi}_2(t)$. For the local noise, this is correct, but the transduced noise \emph{must} have a finite time delay due to the separation of the resonators (thermal baths) and the non-zero travel time of the photons between them: $\delta \hat{x}_1 (t)$ must depend on $\hat{\xi}_1(t)$ and $\hat{\xi}_2(t-\bar{t})$ for an average photon travel time $\bar{t}$. The well-established framework of Eq.~\eqref{QLE} breaks down: it does not contain this time delay. It predicts an immediate response of e.g. resonator 1 when resonator 2 is moved, regardless of the finite photon travel time. Note that the noise transduced by the cavity is first experienced by the other resonator from its own thermal bath (Fig.~\ref{Figure2}(a)).

We introduce the time delay of the transduced noise with respect to the local noise in the autocorrelation of the effective thermal noise experienced by a resonator (e.g. resonator 1), 
\begin{equation}
\begin{aligned}
\langle \hat{\xi}_1^\mathrm{eff}(t) \hat{\xi}_1^\mathrm{eff}(t') + \hat{\xi}_1^\mathrm{eff}(t') \hat{\xi}_1^\mathrm{eff}(t)\rangle/2  
= \left\langle\left(\hat{\xi}_1(t) + M_1(t)*\hat{\xi}_2(t)\right)\left(\hat{\xi}_1(t') + M_1(t')*\hat{\xi}_2(t')\right)\right\rangle& \\ 
\Rightarrow \left\langle\left(\hat{\xi}_1(t) + M_1(t) *\hat{\xi}_2(t-\bar{t})\right)\left(\hat{\xi}_1(t') + M_1(t') *\hat{\xi}_2(t'-\bar{t})\right)\right\rangle&,
\end{aligned}
\end{equation}
where $M(t) = \mathcal{F}^{-1}\lbrace M_1(\omega) \rbrace$ from the inverse Fourier transform, the time delay between the local and transduced noise is $\bar{t}$, and $*$ denotes convolution. We have explicitly introduced the delay time $\bar{t}$ only in the transduced noise term; by property of the convolution we could have distributed the time delay freely between $M_1(t)$ and $\hat{\xi}_2(t)$ without affecting the result. In the frequency domain, using the time-shift property of the Fourier transform, we get a phase shift,

\begin{multline}
\langle \hat{\xi}_1^{\mathrm{eff},'}(\omega) \hat{\xi}_1^{\mathrm{eff},'}(\omega') + \hat{\xi}_1^{\mathrm{eff},'}(\omega') \hat{\xi}_1^{\mathrm{eff},'}(\omega)\rangle/2 = \\ \left\langle\left(\hat{\xi}_1(\omega) + e^{-2i\pi \bar{t}\omega} M_1(\omega)\hat{\xi}_2(\omega)\right)\left(\hat{\xi}_1(\omega') + e^{-2i\pi \bar{t}\omega} M_1(\omega')\hat{\xi}_2(\omega')\right)\right\rangle
\end{multline}
where we have denoted the effective noise with added time delay by $\hat{\xi}_j^{\mathrm{eff},'}$. The frequency range of interest is close to the mechanical frequencies ($\omega \sim \omega_1 \simeq \omega_2$), so we can consider it as a constant phase factor $e^{-2i\pi \bar{t} \omega} \simeq e^{i\phi_1}$. We call this the \emph{optomechanical phase lag} that the transduced noise experiences with respect to the local noise. This modifies Eq.~\eqref{Effectivenoise} to
\begin{equation}
\label{Effectivenoisephase}
\hat{\xi}_1^{\mathrm{eff},'}(\omega) = \hat{\xi}_1+ \alpha_1 e^{i\phi_1} M_1 \hat{\xi}_2, \quad
\hat{\xi}_2^{\mathrm{eff},'}(\omega) = \hat{\xi}_2 + \alpha_2 e^{i\phi_2}M_2 \hat{\xi}_1,
\end{equation}
where we have introduced the amplitude fit factors $\alpha_1$ and $\alpha_2$ to account for imperfect alignment between the two membranes.

Some closer considerations of this time delay and phase lag are warranted. An optomechanically scattered photon traveling the distance between resonators 1 and 2 (\SI{200}{\micro\meter}) takes about \SI{670}{\femto\second}, which should be negligible on the time scale of the mechanical motion, so we would expect the phase lag to be negligibly small as well. However, due to the optical cavity and the fact that $g_{0,j}$ is small, the chance for a scattered photon to directly interact with the other resonator is very small. It is much more likely to exit the cavity without interacting with the other membrane, as $\kappa \gg g_{0,j}$. The photons that \emph{do} interact with the other membrane (i.e. the ones that have not exited the cavity) will thus have an average travel time equal to the lifetime of the cavity $\bar{t} = \tau = 1/\kappa$ . In the regime $\kappa \simeq \omega_j$, this time lag represents a significant fraction of the mechanical period, meaning that the contributions to the effective noise of a resonator can be perfectly out of phase. When that happens, the effective noise term that resonator 1 experiences is reduced due to the coupling to resonator 2 and its thermal bath (and vice-versa). In other words, the local noise and the noise transduced by the optical field from the other resonator interfere. We estimate the optomechanical phase lag for systems from literature (supplemental document Sec.~5) and distinguish interference due to the this effect from other interference mechanisms (supplemental document Sec.~6).

\subsection{Experimental observation of interference}
\begin{figure}[t]
\centering
\includegraphics[width = 0.7\textwidth]{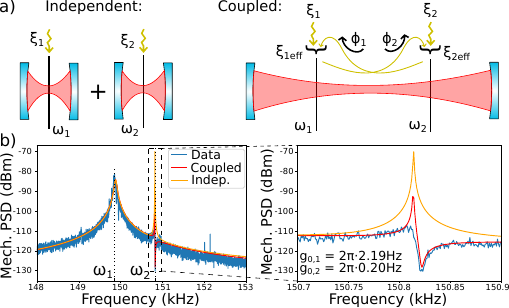}
\caption{a) Schematic of mechanical noise contribution for two independent (left) or cavity-coupled (right) resonators. b) Measured mechanical spectra for two membranes in a single cavity, where one membrane ($\omega_1$) is significantly stronger coupled (more damped) than the other ($\omega_2$). Theory models are fitted for the independent (uncoupled) and the coupled membranes case including optomechanical phase lag.}
\label{Figure2}
\end{figure}

We study the behavior of our optomechanical system by measuring the mechanical power spectral density (PSD) with our homodyne setup. This allows us to test the theory curves obtained with the inclusion of the time delay and the curves obtained for two completely independent membranes (i.e. $G_j$ set to zero while $G_{i\neq j} \neq 0$, for either membrane with the resulting spectra summed), shown in Fig.~\ref{Figure2}(b). As these measurements are in the frequency domain, we shall refer to the optomechanical phase lag rather than the time delay.

The theory curve for the independent resonators (orange, solid line) clearly shows two Lorentzians, one at $\omega_1$ which is broadened due to opomechanical cooling, and one which is less coupled at $\omega_2$, and therefore less broad. The theory curve with the added optomechanical phase lag (red, solid line) follows the other theory curve for most of the frequency domain: because the Lorentzian at $\omega_2$ is narrow, the noise contribution from $\hat{\xi}_2$ to $\hat{\xi}_1^\mathrm{eff}$ is only relevant for a small frequency range around $\omega_2$ (inset). Here, the interference between the noise terms results in a characteristic Fano-lineshape~\cite{Fano1961} in the theory curve where the spectra of the individual mechanical resonators would overlap. 

Comparing both theory curves to the experimental data (blue, solid line), we see a clear drop in the PSD around $\omega_2$, which the theory that includes the optomechanical phase lag describes well, while the model without it does not. Note that the peak of the Fano-lineshape is absent from the experimental data as well, which we attribute this to experimental imperfections. 

\begin{figure}
	\includegraphics[width = 1.0\textwidth]{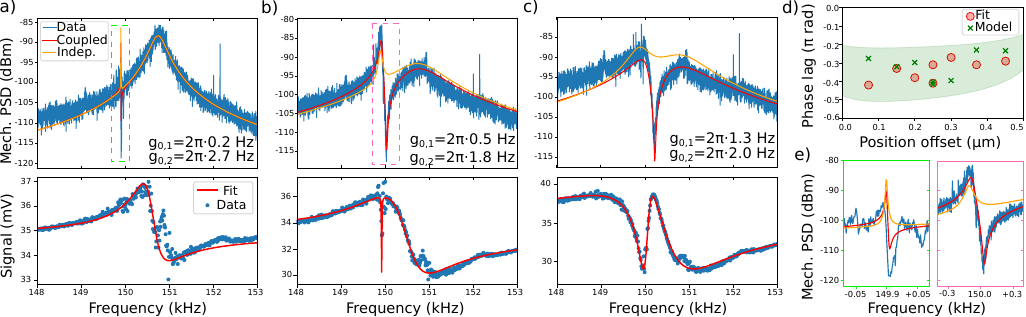}
	\caption{a-c), Top panels:\ Measured PSD for the two-membrane system for various coupling ratios (blue), and the expected behavior with interference (red) and considering independent membranes (orange). The measured coupling rates are shown, their ratios being 0.07, 0.28 and 0.65, respectively. Bottom panels:\ OMIT data (blue dots) and fits (red line) used to extract the optomechanical parameters. d) Extracted phase lag from the fit (red circles) and expected phase lag based on the cavity linewidths (green crosses). The green shaded area shows the phase lag bounds based on the cavity linewidths, see main text. e) Zoom-in on dashed regions of a) and b) showing narrow spectral features at the mechanical frequency of the less-coupled resonator.}
	\label{Couplingsweep}
\end{figure}

To further study how this interference based on the optomechanical phase lag behaves, we adjust the optomechanical coupling rates of the resonators. This changes the frequency range over which the interference is observable, and also its strength. By varying the position of the chip within the cavity, the optical field intensity that each membrane experiences is changed, which allows us to control the optomechanical coupling rate of each of the membranes. We consider four cases, one shown in Fig.~\ref{Figure2}(b) ($g_{0,1} \gg g_{0,2}$), and three in Fig.~\ref{Couplingsweep}(a-c), $g_{0,1} \ll g_{0,2}$, $g_{0,1} < g_{0,2}$ and $g_{0,1} \simeq g_{0,2}$. For the latter three, we also show optomechanically induced transparency (OMIT) measurements and fits~\cite{Weis2010,Agarwal2010,Nielsen2017} (for details see the supplemental document Sec.~7), by which we independently obtain all optomechanical parameters. In these OMIT measurements (Fig.~\ref{Couplingsweep}(a-c) bottom row), we observe an additional feature not captured by our fit. Due to its frequency, it likely stems from the resonator's thermal noise.

In the case where one of the resonators has very weak coupling to the optical field, as shown in Fig.~\ref{Figure2}(b) and Fig.~\ref{Couplingsweep}(a), the PSD of the more strongly-coupled resonator takes the expected Lorentzian form. At the frequency of the weakly-coupled resonator, we observe a consistent dip in the PSD (Figs.~\ref{Figure2}(b) right panel, \ref{Couplingsweep}(e)), where the noise drops $15-20$~\si{\decibel} below the level of the spectrum of the other mode. The optomechanical parameters ($\omega_1 = 2\pi \times 149.89$~\si{\kilo\hertz}, $\omega_2 = 2\pi \times 150.80$~\si{\kilo\hertz}, $\kappa \simeq 2\pi \times 600$~\si{\kilo\hertz}, $\Delta = 2\pi \times 10$~\si{\kilo\hertz}, $g_{0,1} = 2\pi \times 2.2$~\si{\hertz} and $g_{0,2} = 2\pi \times 0.2$~\si{\hertz} for Fig.~\ref{Figure2}(b)) are also obtained through the separate OMIT measurement and fit. 

When one of the resonators is less coupled, but not very weakly, Fig.~\ref{Couplingsweep}(b,e), we see a clear Fano-lineshape in the PSD. If both resonators are approximately equally coupled (cf.\ Fig.~\ref{Couplingsweep}(c)), the measured spectrum exhibits a pronounced anti-resonance~\cite{Rodrigues2009}, clearly signaling destructive interference. There is an additional mode at \SI{147}{\kilo\hertz} that is not included in the fits in Fig.~\ref{Couplingsweep}. Combined, these measurements show that our model with phase lag consistently describes the experimental data much better (\SI{20}{\deci\bel}, a factor 100 difference) than the theory without the phase lag. 

The most important factor governing the phase lag is the cavity linewidth $\kappa$. As we change the position of the chip in the cavity, $\kappa$ changes due to scattering and misalignment losses. We plot the expected phase lag as a function of the chip position in Fig.~\ref{Couplingsweep}(d), by way of the fits (red circles) to the spectra of Figs.~\ref{Figure2},~\ref{Couplingsweep} and~\ref{Figure4}, and the calculated values (green crosses) based on the average $\kappa$ measured directly before and after each experiment. Unfortunately, $\kappa$ is the main source of uncertainty in the theory curves, as it has a significant uncertainty from the OMIT fits and a spread ($275-600$~\si{\kilo\hertz}) when measured directly from a laser wavelength scan before and after OMIT and PSD measurements. This is likely due to our imperfect stabilization of the laser to the cavity frequency, smeared out by the averaging. To illustrate, we have calculated the expected phase lag for a spread of $\kappa = 275-600$~\si{\kilo\hertz} (green shaded area) all observed from measurements at the same wavelength and chip position. We have calculated the expected linewidths based on this spread using a model of a Fabry-Pérot cavity with lossy membranes (supplemental document Sec.~8). There is reasonable agreement between the fitted and calculated values, and all values fall well within the band based on the spread in $\kappa$.

\subsection{Cooperativity competition}
Independently of the optomechanical phase lag, we predict that if two mechanical resonators are coupled to the same optical field, the effective mechanical dissipation of one does not only depend on its local environment but also on the optomechanical cooperativity of the other resonator. We refer to this effect as \textit{cooperativity competition} (for details see the supplemental document, Sec.~9). From the solution to Eq.~\eqref{QLE}, we can rewrite the position fluctuations in terms of the effective susceptibility $\chi_i^\mathrm{eff} (\omega)$. We can further define the effective mechanical linewidths, which reduce to the simple expressions
\begin{equation}
\label{CClinewidths}
\gamma_1^\mathrm{eff} \approx \gamma_1\left( 1 + \frac{C_1}{C_2} \right), \quad\quad
\gamma_2^\mathrm{eff} \approx \gamma_2\left( 1 + \frac{C_2}{C_1} \right). 
\end{equation}
Here we assume identical mechanical frequencies and optimal cooling, $\Delta = \omega_1 = \omega_2$, side-band resolution, $\kappa \lesssim \omega_j$ and  large optomechanical cooperativities, $C_j = 2G_j^2/(\kappa \gamma_j) \gg 1$. These equations describe how the effective mechanical dissipation of one resonator is reduced with respect to those of two independent modes, where $\gamma_j^{\rm eff} \simeq \gamma_j (1+C_j)$ ($j=1,2$)~\cite{WilsonRae2007,Marquardt2007,Genes2008}. While Eq.~\eqref{CClinewidths} describes a simplified model, for our experiments we use the full model (see the supplemental document, Sec.~9) to obtain $\gamma_j^\mathrm{eff}$. Although both the optomechanical phase lag and the cooperativity competition originate from cavity-mediated coupling between the mechanical resonators, they are essentially different effects with their own characteristics, embodied by noise cancellation and competition in dissipation dynamics respectively.

\begin{figure}
\centering
\includegraphics[width = 0.7\textwidth]{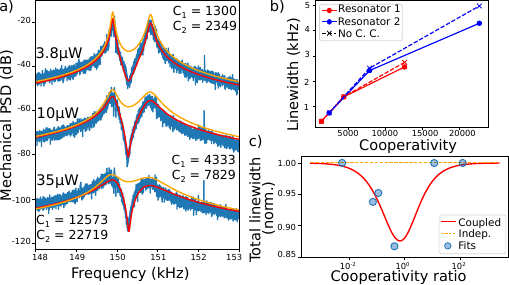}
\caption{a) PSD for a fixed $g_{0,1}/g_{0,2} = 0.74$ and various powers (blue), with theory fits for the coupled-resonator (red) and independent-resonator models (orange). b) Fitted mechanical linewidths as a function of cooperativity. Solid lines include the effect of cooperativity competition while dashed lines do not. c) Normalized total (sum) linewidths for various coupling ratios (blue), showing a decrease due to cooperativity competition.}
\label{Figure4}
\end{figure}

To observe cooperativity competition in our system, we vary cooperativities $C_1$ and $C_2$ (see Eq.~\eqref{CClinewidths}) by changing the coupling ratio $g_{0,1}/g_{0,2}$ or by changing the optical power. With $g_{0,1}/g_{0,2} = 0.74$ to keep the effect of the interference on the shape of our PSD constant, we measure at different powers, Fig.~\ref{Figure4}(a) (blue). The optomechanical parameters are determined as before, which we then use to fit our coupled (red) and independent (orange) models. The cooperativity competition manifests itself as a change in linewidth of the two resonances, which is difficult to gauge from the shape of the PSD, as it is dominated by the interference. Therefore we have plotted the fitted linewidths in terms of the cooperativity in Fig.~\ref{Figure4}(b) for both the coupled case (solid curves), which contains both the interference and the cooperativity competition, and the independent case (dashed curve) which contains neither. This shows an appreciable reduction in linewidth for higher cooperativities as predicted. 

We can further corroborate cooperativity competition by analyzing the fitted linewidths for various coupling ratios, shown in Fig.~\ref{Figure4}(c). We compare the total linewidth (sum of both linewidths), and expect a straight line as a function of $C_1/C_2$ in the independent case (orange, dashed), while cooperativity competition predicts a cooperativity-ratio-dependent reduction of the total linewidth (red, solid). The reduction is maximal when the cooperativities are approximately equal where the competition is most intense, and the curve is symmetric around $C_1/C_2 = 1$, which can be seen by switching the labels $1,~2$ of the resonators. The fitted linewidths are normalized to account for the cooling efficiency by rescaling the total linewidth by the maximum reduction expected due to cooperativity competition for the fitted $\kappa$, $\Delta$ and cooperativities of each data point. The results match with the expected decrease associated with the cooperativity competition as a function of $C_1/C_2$. This shows the effective optomechanical coupling leading to a competition on the mechanical dissipation of the resonators.

\section{Conclusion}
We have introduced an optomechanical phase lag between the local and cavity-transduced thermal noises of the two resonators, originating from the time-delay of noise transduced via the cavity. We have observed interference stemming from this phase lag by measuring the mechanical power spectral density of a double-membrane device. The interference coherently cancels mechanical noise of the two resonators where their (broadened) frequency spectra overlap, leading to a \SI{20}{\decibel} decrease in mechanical noise. This could create an interesting new method of controllably reducing unwanted mechanical noise by introducing a second resonator, which would allow cancellation of mechanical noise in a specific frequency range (supplemental document, Sec.~10). 

In addition, we have proposed and experimentally verified another new collective effect in the same system, where the effective susceptibility of the coupled resonators causes a competition on the mechanical dissipation. The dissipation rates of two mechanical resonators can get significantly reduced when their optomechanical cooperativities are comparable. This novel collective effect paves the way for long-range control of phonon dynamics~\cite{Xuereb2014} and the results of this work can be applied directly to multi-resonator ($N>2$) optomechanical systems, where we expect more prominent and even richer collective effects.

\begin{backmatter}
\bmsection{Funding}
This work is supported by the the European Research Council (ERC StG Strong-Q, 676842), and by the Netherlands Organization for Scientific Research (NWO/OCW), as part of the Frontiers of Nanoscience program, as well as through Vidi (680-47-541/994) and Vrij Programma (680-92-18-04) grants.

\bmsection{Acknowledgements}
We thank Jo\~{a}o P.\ Moura and Klara Knupfer for their work in designing the cavity-membrane setup, and Girish S.\ Agarwal and David Vitali for helpful discussions. 

\bmsection{Disclosures}
The authors declare no conflicts of interest.

\bmsection{Data availability}
All data, measurement and analysis scripts in this work are available at\newline \href{https://doi.org/10.5281/zenodo.5782970}{https://doi.org/10.5281/zenodo.5782970}.
\end{backmatter}

\setcounter{figure}{0}
\renewcommand{\thefigure}{S\arabic{figure}}
\setcounter{equation}{0}
\renewcommand{\theequation}{S\arabic{equation}}

\clearpage

\section{Supplemental}
This supplemental document contains the derivation of the equations of motion for a set of two mechanical resonators coupled to a single optical mode, as well as the derivation of the mechanical power spectral density observable from this system in a homodyne detection setup. We also include the full analytical derivation for cooperativity competition. We provide supporting material and calculations for the experiments and conclusions drawn in the main text, i.e. a quantitative estimate of the rate of the scattering-rescattering process (effective mechanics-mechanics beam splitter), the effective temperature of our mechanical resonators due to optomechanical cooling, a comparison of optomechanical phase lag expected in other systems in literature, qualitative and quantitative arguments that allow exclusion of other interference mechanisms as explanations of the observations in the main text, a classical model of a Fabry-Pérot cavity that allows us to model optical and optomechanical parameters, a derivation of multi-mode optomechanically induced transparency (OMIT) and a proposal for application of the optomechanical phase lag to a sensor.

\subsection{Equations of motion for the coupled resonator system}\label{QLEderivation}

We consider a system where two mechanical modes (each the fundamental mechanical mode of a membrane) are coupled to an optical cavity via radiation pressure. There is no direct coupling between the two mechanical modes, but they are indirectly coupled by the mediation of light. The Hamiltonian of the system is given by
\begin{equation}\label{Hamiltonian}
\frac{H}{\hbar} = \omega_c \hat{a}^{\dag} \hat{a} + \!\! \sum_{j=1,2} \bigg[  \frac{\omega_j}{2} (\hat{x}_j^2 + \hat{p}_j^2) - g_{0,j} \hat{a}^{\dag} \hat{a} \hat{x}_j  \bigg] + i E \big(\hat{a}^{\dag} e^{-i \omega_\ell t}  - {\rm H.c.} \big),
\end{equation}
where $\hat{a}$ ($\hat{a}^{\dag}$) is the annihilation (creation) operator of the cavity field, $\hat{x}_j$ and $\hat{p}_j$ are, respectively, the dimensionless position and momentum of the $j$\textsuperscript{th} ($j\,{=}\,1,2$) mechanical resonator, and thus we have $[\hat{a}, \hat{a}^{\dag}]=1$ and $[\hat{x}_j, \hat{p}_j]=i$. The resonance frequencies $ \omega_c$, $ \omega_j$ are of the cavity and the $j$\textsuperscript{th} mechanical resonator, respectively, and $g_{0,j}$ is the single-photon optomechanical coupling rate related to the $j$\textsuperscript{th} mechanical resonator. The last term in the Hamiltonian denotes the laser driving for the cavity, where $E=\!\sqrt{P_\ell \kappa_\mathrm{e}/\hbar \omega_\ell}$ is the coupling between the cavity with external decay rate $\kappa_\mathrm{e}$ and the driving laser with frequency $\omega_\ell$ and power $P_\ell$.

In the frame rotating at the drive frequency $\omega_\ell$ and by including input noises and dissipation of the system, we obtain the following quantum Langevin equations (QLEs), which govern the system dynamics
\begin{equation}\label{QLE1}
\begin{split}
\dot{\hat{x}}_j&= \omega_j \hat{p}_j,   \\
\dot{\hat{p}}_j&= - \omega_j \hat{x}_j - \gamma_j \hat{p}_j + g_{0,j} \hat{a}^{\dag} \hat{a} + \hat{\xi}_j,  \qquad (j=1,2)  \\
\dot{\hat{a}}&= - (i \Delta_0 + \kappa/2) \hat{a} + i \sum_{j=1,2} g_{0,j} \hat{x}_j  \hat{a} + E + \sqrt{ \kappa} \hat{a}^{\rm in} ,  \\
\end{split}
\end{equation}
where $\Delta_{0} \,\,{=}\,\, \omega_{c} \,{-}\, \omega_\ell$, $\kappa$ is the total cavity decay rate ($\kappa > \kappa_e$), $\gamma_j$ is the mechanical damping rate, $\hat{a}^{\rm in} $ denotes vacuum input noise for the cavity, whose mean value is zero and the only nonzero correlation is 
\begin{equation}\label{Opticsnoise}
\langle \hat{a}^{\rm in} (t) \, \hat{a}^{\rm in, \dag}(t')\rangle = \delta(t-t').
\end{equation} 
Here, $\hat{\xi}_j$ is the Langevin force operator, which accounts for the Brownian motion of the $j$\textsuperscript{th} mechanical resonator and is auto-correlated as 
\begin{equation}\label{Mechanicsnoise}
\langle \hat{\xi}_j(t)\hat{\xi}_j(t')+\hat{\xi}_j(t') \hat{\xi}_j(t) \rangle/2  \simeq \gamma_j (2 \bar{n}_j+1) \delta(t-t'), 
\end{equation}
where a Markovian approximation has been made. This is valid for a large mechanical quality factors $Q_j= \omega_j/\gamma_j \,\, {\gg}\, 1$~\cite{Benguria1981,Giovannetti2001}, and $\bar{n}_j \simeq \frac{k_B T}{\hbar \omega_j}$ is the mean thermal phonon number in the high temperature limit, with $k_B$ the Boltzmann constant and $T$ the environmental temperature.

In the experiment, the cavity is strongly driven which leads to a large amplitude of the cavity field $|\langle \hat{a} \rangle| \gg 1$. This allows us to linearize the system dynamics around the semi-classical averages by writing any operator as $\hat{O}=\langle \hat{O} \rangle +\delta \hat{O}$ ($\hat{O}\, {=}\, \hat{a},\hat{x}_j,\hat{p}_j$) and neglecting second-order fluctuation terms. We obtain the linearized QLEs for the quantum fluctuations $(\delta \hat{x}_j, \delta \hat{p}_j, \delta \hat{a})$
\begin{equation}\label{QLE2}
\begin{split}
\delta \dot{\hat{x}}_j&= \omega_j \delta \hat{p}_j,   \\
\delta \dot{\hat{p}}_j&= - \omega_j \delta \hat{x}_j - \gamma_j \delta \hat{p}_j + G^*_j \delta \hat{a} + G_j \delta \hat{a}^{\dag}  + \hat{\xi}_j,   \\
\delta \dot{\hat{a}}&= -(i \Delta + \kappa/2 ) \delta \hat{a} + i \sum_{j=1,2} G_j \delta \hat{x}_j  + \sqrt{ \kappa} \hat{a}^{\rm in} 
\end{split}
\end{equation}
where (complex) $G_j \,{=}\, g_{0,j} \langle \hat{a} \rangle$ is the effective optomechanical coupling rate, $\langle \hat{a} \rangle = \frac{E}{\kappa/2+ i \Delta} $, and $\Delta \,{=}\, \Delta_0 \,{-} \sum_{j} \frac{g_{0,j}^2}{ \omega_j} |\langle \hat{a} \rangle|^2$ is the effective detuning.

By taking the Fourier transform of each equation in \eqref{QLE2} and solving {\it separately} the two equations for each mode in the frequency domain, we obtain the following solutions
\begin{eqnarray}
&&\delta \hat{p}_j = -i \frac{ \omega}{\omega_j} \delta \hat{x}_j , \label{p_j}  \\
&&\delta \hat{x}_j = \, \chi_j (\omega) \Big[G_j^* \delta \hat{a} + G_j \delta \hat{a}^{\dag} + \hat{\xi}_j \Big],  \label{x_j} \\  
&&\delta \hat{a} = \chi_c (\omega) \Bigg( \sum_{j=1,2} i G_j \delta \hat{x}_j + \!\sqrt{\kappa} \hat{a}^{\rm in}  \Bigg),    \label{aa} \\  
&&\delta \hat{a}^{\dag} = \chi_c^* (-\omega) \Bigg( \sum_{j=1,2} -i G_j^* \delta \hat{x}_j + \!\sqrt{\kappa} \hat{a}^{\rm in, \dag}  \Bigg),    \label{ad} 
\end{eqnarray}
where we have introduced the natural susceptibility of the mechanical resonators, $\chi_j (\omega)$, and of the cavity field, $\chi_c (\omega)$, given by
\begin{eqnarray}
&&\chi_j (\omega) =  \frac{\omega_j}{\omega_j^2 - \omega^2 - i \gamma_j \omega},   \label{chi_j} \\
&&\chi_c (\omega) =  \frac{1}{ \kappa/2 + i (\Delta - \omega)  \label{chi_c} }, \\ 
&&\chi^*_c (-\omega) =  \frac{1}{ \kappa/2 - i (\Delta + \omega)}.
\end{eqnarray}

Solving \eqref{x_j} for $\delta \hat{x}_2$ (i.e., taking $j=2$) and \eqref{aa} and~\eqref{ad} for $\delta \hat{a}$ and $\delta \hat{a}^{\dag}$, and inserting their solutions into \eqref{x_j} of $\delta \hat{x}_1$ (i.e., taking $j=1$), we obtain

\begin{equation} \label{eqx1s}
\delta \hat{x}_1  =  \chi_1^{\rm eff} (\omega) \left\lbrace \frac{ \splitfrac{-\big[G_1^* G_2 \chi_c (\omega) - G_1 G_2^* \chi_c^* (-\omega) \big] \, \chi_2 (\omega)  \, \hat{\xi}_2}{ +  \sqrt{ \kappa} \big[ (c_1 + iG_1^* \chi_c(\omega)) \hat{a}^{\rm in} + (c_2 + iG_1 \chi_c^*(-\omega))\hat{a}^{\rm in, \dag}\big] }} {i +  G_2^* G_2 \, \chi_2 (\omega)  \big[ \chi_c (\omega) -  \chi_c^* (-\omega) \big]} + \hat{\xi}_1 \right\rbrace ,
\end{equation}
with 
\begin{equation}
\begin{aligned}
c_1 &= \chi_2 \chi_c(\omega) \chi_c^*(-\omega) G_2^* \left(G_1 G_2^* - G_1^* G_2\right) \\
c_2 &= \chi_2 \chi_c(\omega) \chi_c^*(-\omega) G_2 \left(G_1 G_2^* - G_1^* G_2\right) 
\end{aligned}
\end{equation}
which is fully solved and a function of only input noise operators $\big( \hat{\xi}_1, \hat{\xi}_2, \hat{a}^{\rm in} , \hat{a}^{\rm in, \dag} \big)$. \eqref{eqx1s} recognizes three noise sources for the first mechanical resonator: the noise from its own thermal bath, the optomechanical back-action noise from the cavity field, and the thermal noise from the second mechanical resonator transduced through the cavity field. 
We have defined the effective susceptibility
\begin{equation}\label{chi1effs}
\chi_1^{\rm eff} (\omega) = \Bigg[ \frac{1}{\chi_1 (\omega)}  +  \frac{G_1^* G_1 }{\frac{i}{\chi_c (\omega) - \chi_c^* (-\omega)} + G_2^* G_2 \, \chi_2 (\omega)} + d_1  \Bigg]^{-1}
\end{equation}
with
\begin{equation}
d_1 = \frac{\chi_2(\omega) \chi_c (\omega) \chi_c^* (-\omega) \left(G_1 G_2^* - G_1^* G_2 \right)^2}{i G_2^* G_2 \chi_2 (\omega) \left(\chi_c (\omega) - \chi_c^* (-\omega)\right) -1},
\end{equation}
which identifies the coupling to the cavity field and the {\it indirect} coupling to the second mechanical resonator mediated by the light. Taking $G_2=0$, i.e.\ the second mechanical resonator is decoupled from the cavity field, we obtain
\begin{equation}
\chi_1^{\rm eff} (\omega) =  \Bigg[ \frac{1}{\chi_1 (\omega)}  -i G_1^* G_1 \big[ \chi_c (\omega) - \chi_c^* (-\omega) \big]  \Bigg]^{-1},
\end{equation}
which is exactly the effective mechanical susceptibility provided in Ref.~\cite{Genes2008} (note:\ their definitions of $G$ and $\kappa$ are slightly different from ours), which studied a single mechanical resonator coupled to cavity field. Owing to the symmetry of the two mechanical resonators, we therefore get the effective susceptibility of the second mechanical resonator
\begin{equation}
\begin{aligned} \label{chi2eff}
\chi_2^{\rm eff} (\omega) &= \Bigg[ \frac{1}{\chi_2 (\omega)}  +  \frac{G_2^* G_2 }{\frac{i}{\chi_c (\omega) - \chi_c^* (-\omega)} + G_1^* G_1 \, \chi_1 (\omega)} + d_2   \Bigg]^{-1}\\
d_2 &= \frac{\chi_1(\omega) \chi_c (\omega) \chi_c^* (-\omega) \left(G_1 G_2^* - G_1^* G_2 \right)^2}{i G_1^* G_1 \chi_1(\omega) \left(\chi_c (\omega) - \chi_c^* (-\omega)\right) -1}.
\end{aligned}
\end{equation}
By neglecting the optical input noise, which is small compared to the room-temperature thermal noise in our system, we can express the position fluctuation in terms of an effective susceptibility and effective noise%~\cite{Boyanovsky2017}

\begin{equation}\label{xxx1}
\delta \hat{x}_1 = \chi_1^{\rm eff} (\omega) \hat{\xi}_1^\mathrm{eff} = \chi_1^{\rm eff} \left( \hat{\xi}_1 + M_1 \hat{\xi}_2 \right)
\end{equation}
with 
\begin{equation}
M_1 = \frac{i\chi_2(\omega) \left( G_1^* G_2 \chi_c (\omega) - G_1 G_2^* \chi_c^*(-\omega) \right)}{1 - i G_2 G_2^* \chi_2(\omega) (\chi_c(\omega) - \chi_c^*(-\omega))}
\end{equation}
describing the transduction of the mechanical noise from resonator $2\rightarrow 1$ through the cavity field, as in the main text. Note that we have considered the general situation of complex couplings $G_j$ in $M_1$. Using the same approach, we can write the position fluctuation $\delta \hat{x}_2$ in a similar form as \eqref{xxx1}.

\subsection{Homodyne detection of mechanical fluctuations}\label{Homodynespectrum}
We derive the expected power spectral density (PSD) that we would detect in our homodyne detection setup based on the fluctuations of the mechanical operators, $\delta \hat{x}_j$ and $\delta \hat{p}_j$ and the optical field $\delta \hat{a}$. We start from \eqref{p_j}-\eqref{ad}, which we solve for the fluctuations $\delta \hat{x}_j$, $\delta \hat{p}_j$ and $\delta \hat{a}$ in terms of the noise operators $\hat{\xi}_j$ and $\hat{a}^\mathrm{in}$.

We use a homodyne detection setup that is sensitive to the optical field output from our cavity, so we are interested in finding the PSD of that field rather than the PSD of the mechanical fluctuations. Using input-output theory, we can obtain the output field as
\begin{equation}
\delta \hat{a}^\mathrm{out} = \sqrt{\kappa_\mathrm{e}} \delta \hat{a} - \hat{a}^\mathrm{in}.
\end{equation}
Using the results from the previous section, we have
\begin{equation}\label{Fieldterm}
\begin{aligned}
\delta \hat{a} &= \chi_c \left( i G_1 \chi_1^\mathrm{eff} (\hat{\xi}_1 + M_1\hat{\xi}_2) + i G_2 \chi_2^\mathrm{eff} (\hat{\xi}_2 + M_2 \hat{\xi}_1)\! + \! \sqrt{\kappa} \hat{a}^\mathrm{in} \right) \\
\delta \hat{a}^\dagger \!\! &= \chi_c^*\left(- i G_1^* \chi_1^\mathrm{eff} (\hat{\xi}_1\! +\! M_1\hat{\xi}_2) - i G_2^* \chi_2^\mathrm{eff} (\hat{\xi}_2\! +\! M_2 \hat{\xi}_1)\! + \!\! \sqrt{\kappa} \hat{a}^\mathrm{in,\dagger} \right)
\end{aligned}
\end{equation}
To consider the fact that we have a homodyne detection where our local oscillator might have a phase offset $\phi$ with regards to the signal returned from the cavity, we observe a general quadrature $\delta \hat{z}^\mathrm{out}$, given by
\begin{equation}
\delta \hat{z}^\mathrm{out} = \frac{1}{\sqrt{2}} \left( \delta \hat{a}^\mathrm{out} e^{-i\phi} + \delta \hat{a}^{\mathrm{out},\dagger} e^{i \phi}\right).
\end{equation}
We can reorganize the expression for $\delta \hat{z}^\mathrm{out}$ in the form of coefficients of these noise terms, 
\begin{equation}
\delta \hat{z}^\mathrm{out} = \delta \hat{z}^\mathrm{out}_a \hat{\xi}_1 + \delta \hat{z}^\mathrm{out}_b \hat{\xi}_2 + \delta \hat{z}^\mathrm{out}_c \hat{a}^\mathrm{in} + \delta \hat{z}^\mathrm{out}_d \hat{a}^{\mathrm{in},\dagger}.
\end{equation}
The quantity of interest is the spectrum of these fluctuations of the generalized output quadrature, which can be obtained by calculating
\begin{multline}
S_{z}(\omega) = \\ \frac{1}{2\pi}\int_\infty^\infty \frac{1}{2}\langle \delta \hat{z}^\mathrm{out}(\omega) \delta \hat{z}^\mathrm{out} (\omega') + \delta \hat{z}^\mathrm{out} (\omega') \delta \hat{z}^\mathrm{out} (\omega)\rangle e^{-i(\omega + \omega')t}\mathrm{d}\omega'.\label{PSDeq}
\end{multline}
When we rewrite this equation in terms of the coefficients of the noise operators, we obtain terms containing the correlations of the noise, which we already know (\eqref{Opticsnoise} and \eqref{Mechanicsnoise}). From there it is straightforward, if tedious, to calculate the PSD from only the cavity and mechanical parameters.

\subsection{Quantitative estimation of scattering-rescattering rate}
\begin{figure}
\includegraphics[width = 1.0\textwidth]{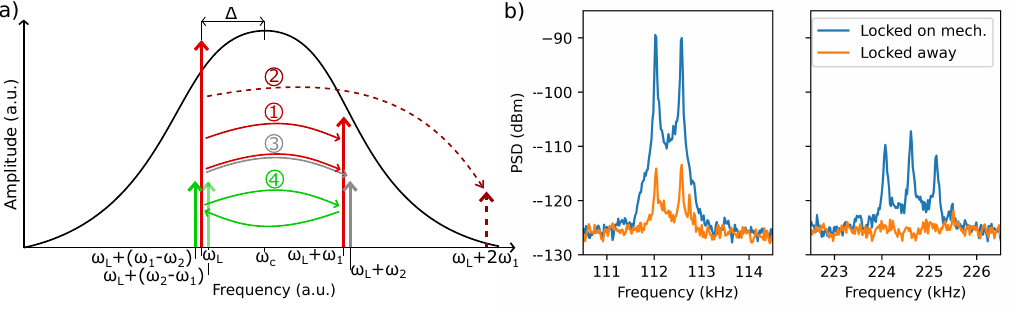}
\caption{a): Schematic of (some) processes in the optomechanical system: (1) First-order Anti-Stokes scattering from laser ($\omega_\mathrm{\ell}$) to mechanical sideband ($\omega_\mathrm{\ell} + \omega_1$). (2) Second-order Anti-Stokes scattering to $\omega_\mathrm{\ell} + 2\omega_1$. (3) Simultaneous first-order Anti-Stokes scattering to both resonators $\omega_1$ and $\omega_2$. (4) Subsequent scattering (Anti-Stokes then Stokes) to $\omega_1$ and $\omega_2$, and to $\omega_2$ and $\omega_1$. b): Mechanical PSD showing first (left) and second (right) order scattering processes from a different set of resonators with lower resonance frequencies, for comparable cavity and laser drive parameters. We clearly resolve the two first-order processes at $\omega_1$ and $\omega_2$. While the second order peaks at $2\omega_1$, $2\omega_2$ and $\omega_1 + \omega_2$ are much smaller than the first, they are still detectable.}
\label{FigS11}
\end{figure}

The scattering-rescattering process that forms the effective mechanics-mechanics beam-splitter interaction between the resonators is a second-order optical effect, but it is linear in the mechanical operators. To illustrate its relation to other optomechanical processes happening in our system, we have sketched it in Fig.~\ref{FigS11}(a). The process of interest is (4), the scattering and subsequent rescattering of a single photon to/from a resonator. 

To estimate the rate of this interaction, and corroborate that this can be of the order of the mechanical noise, we make a quantitative estimate of the size of this effect. We can calculate the strength of the first-order scattered fields with respect to the laser-driven cavity field by using Eqs.~(60) and (61) of~\cite{Aspelmeyer2014},
\begin{equation}
\begin{aligned}
a &\simeq a_0 + a_1 \\
a_0 &= a_\mathrm{in} \frac{\sqrt{\kappa_\mathrm{e}}}{i\Delta + \kappa/2} \\
a_1 &= \frac{g_0 x_0}{2 x_\mathrm{ZPF}}a_0 \left( \frac{1}{i(\Delta-\omega_j) + \kappa/2} - \frac{1}{i(\Delta+\omega_j) + \kappa/2}\right)  \\
&= a_\mathrm{aS} + a_\mathrm{S},
\end{aligned}
\end{equation}
where $a$ is the total cavity field, split in the laser carrier field $a_0$ and the first-order scattered field $a_1$ (containing both anti-Stokes and Stokes sidebands, $a_\mathrm{aS} + a_\mathrm{S}$). Here we neglect the time-dependence of these fields, as we are only interested in their amplitude. We have mechanical motional amplitude $x_0$, zero-point motion $x_\mathrm{ZPF}$ and optical input field $a_\mathrm{in} \sqrt{\kappa_\mathrm{e}}$ which is simply the input field (laser) we have previously denoted by $E = \sqrt{P_\ell \kappa_\mathrm{e}/\hbar\omega_\ell}$. We can calculate the zero-point motion via $x_\mathrm{ZPF} = \sqrt{\hbar/(2m_\mathrm{eff}\omega_j)}$ for $m_\mathrm{eff} \approx 29$~\si{\nano\gram} the effective mass of the fundamental mode obtained from a COMSOL model (cf. Fig.~\ref{FigureS5}), taking into account the correct normalization~\cite{Hauer2013}. Similarly, by utilizing the equipartition theorem we estimate the motional amplitude $x_0 \approx 7$~\si{\pico\meter} at room temperature.

For the parameters of Table~\ref{Table1}, but simplifying $\omega_1 = \omega_2 = 150$~\si{\kilo\hertz} and $g_{0,1} = g_{0,2} = 2\pi\times 1.6$~\si{\hertz}, we calculate a ratio $a_\mathrm{S}/a_0 = 0.017$, $a_\mathrm{aS}/a_0 = 0.020$. By treating the first-order scattered fields each as a new 'main' field, we can repeat this and calculate the scattered-rescattered fields $a_\mathrm{2s} = (a_\mathrm{aS})_\mathrm{S} + (a_\mathrm{S})_\mathrm{aS}$ with $(a_\mathrm{aS})_\mathrm{S}$ the Stokes-rescattered (2nd order) sideband of the (1st order) anti-Stokes field. The ratio of the (sum of) these fields with respect to the original cavity field is $a_\mathrm{2s}/a_0 = 9.2\times 10^{-4}$. This is much weaker than the main cavity (laser driven) field. Similarly, we can calculate the amplitude ratio of the other second-order sidebands, $(a_\mathrm{aS})_\mathrm{aS}$ and $(a_\mathrm{S})_\mathrm{S}$ with the original field. We get a ratio of $4.4\times 10^{-4}$.

As the photons that go through the scattering-rescattering process end up around the laser frequency, $\omega_\mathrm{\ell} \pm (\omega_1 - \omega_2)$, we cannot separate them from low-frequency noise in our homodyne detection. We can however detect other second order sidebands (cf.\ (2) in Fig.~\ref{FigS11}(a)) that should have a comparable amplitude based on the calculation above. For a different set of resonators (lower frequency) but with comparable cavity parameters and driving power, we can clearly detect both the first-order scattering process (Fig.~\ref{FigS11}(b)) at $\omega_{1,2}$ and the second-order scattering process (right) at $2\omega_1$, $2\omega_2$ and $\omega_1 + \omega_2$. This is different from first-order scattering from the second mode of the mechanical resonators, which happens around \SI{240}{\kilo\hertz}. When we lock the laser frequency (far) away from the center of the cavity resonance, the signal from the mechanics becomes very weak (orange curve in Fig.~\ref{FigS11}(b)) and we can no longer see the second-order scattering process. This power difference between the detected first- and second-order sidebands corresponds well with the calculations above, confirming our estimation of the strength of the scattering-rescattering process.

To see if this process is also a reasonable explanation for the transfer and subsequent cancellation of thermal noise of our two resonators, we have to compare the rate of photons going through this scattering-rescattering process with the thermal phonon diffusion rate of our resonators. If the rate of photons going through this process is similar to the rate of thermal phonons, it is reasonable for the phonons from resonator 2 to be transferred to resonator 1 (or vice versa), causing cancellation of their mechanical noise (if they are perfectly out of phase). To calculate this, we must multiply the photon number $n = \langle a_\mathrm{2s}^* a_\mathrm{2s}\rangle$ with the cavity decay rate $\kappa$ to gain the number of photons per second. However, the amplitude of the scattered fields depends on the position fluctuations $x_0$, which depend on the effective mode temperature. By optomechanical cooling, we drastically reduce the mode temperature, and thus the sideband strength. We have calculated the steady-state effective temperature of the resonators as described in the next section (approximately \SI{0.1}{\kelvin}) which reduces our motional amplitude to \SI{128}{\femto\meter}. This amounts to approximately $7.6 \times 10^{6}$ photons per second going through the scattering-rescattering process to exchange mechanical noise between the resonators. By comparing this with the average thermal phonon number per second for one of the mechanical resonators, about $4.2 \times 10^{6}$ (calculated for a \SI{150}{\kilo\hertz} resonator at \SI{0.1}{\kelvin} effective mode temperature with a \SI{3}{\kilo\hertz} effective linewidth conforming to the optomechanical cooling achieved for a system with the parameters of Table~\ref{Table1}), we can clearly see that a significant fraction of thermal phonons can be transduced between the resonators (and therefore experience this phase delay) within the thermal decoherence time of that resonator.

\begin{table}
\begin{center}
\begin{tabular}{ll}\hline
Parameter & Value \\
\hline
$\omega_1$ & $2\pi \times 149.90$ \si{\kilo\hertz} \\
$\omega_2$ & $2\pi \times 150.83$ \si{\kilo\hertz} \\
$\kappa$ & $2\pi \times 320$ \si{\kilo\hertz} \\
$\Delta$ & $2\pi \times 23.5$ \si{\kilo\hertz} \\
$g_{0,1}$ & $2\pi \times 1.3$ \si{\hertz} \\
$g_{0,2}$ & $2\pi \times 2.0$ \si{\hertz} \\
$\phi_1 = -\phi_2$ & $-0.27$ \\
$\alpha_1$ & 1.4 \\
$\alpha_2$ & 1.7 \\
\hline
\end{tabular}
\end{center}
\caption{Optomechanical parameters used in Fig. 3c (main text) and Fig.~\ref{FigureS6} for simulation of the PSD.}
\label{Table1}
\end{table}

\subsection{Effective temperature of mechanical resonators}
\label{Effectivetemperature}
The autocorrelation of the mechanical noise of the resonators, \eqref{Mechanicsnoise}, contains the mechanical linewidth $\gamma_j$ and mean thermal phonon number of the resonator, $\bar{n}_j$. Both of these parameters are affected by the optomechanical cooling, and we use the effective (optomechanically broadened) $\gamma_j^\mathrm{eff}$ and $\bar{n}_j^\mathrm{eff}$ \cite{WilsonRae2007,Marquardt2007,Genes2008} to fit our experimental data. To obtain these values, we adopt an approach solving the Lyapunov equation \cite{Parks1981}. We start from the QLEs for the fluctuations of our system, \eqref{QLE2}, and we write it in a matrix form such that $\dot{u}(t) = A u(t) + n(t)$, where $u(t)$ is the vector of our system coordinates, $A$ is the drift matrix and $n(t)$ contains only the noise terms. We get
\begin{equation}
\begin{pmatrix}
\delta \dot{\hat{x}}_1 \\
\delta \dot{\hat{p}}_1 \\
\delta \dot{\hat{x}}_2 \\
\delta \dot{\hat{p}}_2 \\
\delta \dot{\hat{a}} \\
\delta \dot{\hat{a}}^\dagger
\end{pmatrix}\!
=\!
\begin{pmatrix}
0 & \omega_1 & 0 & 0 & 0 & 0 \\
-\omega_1 & -\gamma_1 & 0 & 0 & G_1^* & G_1 \\
0 & 0 & 0 & \omega_2 & 0 & 0 \\
0 & 0 & -\omega_2 & -\gamma_2 & G_2^* & G_2 \\
iG_1 & 0 & iG_2 & 0 & -(i\Delta + \kappa/2) & 0 \\
-iG_1^* & 0 & -iG_2^* & 0 & 0 & -(-i\Delta + \kappa/2)
\end{pmatrix}\!\!\!
\begin{pmatrix}
\delta \hat{x}_1 \\
\delta \hat{p}_1 \\
\delta \hat{x}_2 \\
\delta \hat{p}_2 \\
\delta \hat{a} \\
\delta \hat{a}^\dagger
\end{pmatrix}
+
\begin{pmatrix}
0 \\
\xi_1 \\
0 \\
\xi_2 \\
\sqrt{\kappa} \hat{a}^\mathrm{in} \\ 
\sqrt{\kappa} \hat{a}^{\mathrm{in},\dagger}
\end{pmatrix}.
\end{equation}
Then, we can define the diffusion matrix $D$ in terms of the entries of our noise vector, 
\begin{equation}
\frac{1}{2}\langle n_i(t) n_j(t') + n_j(t') n_i(t) \rangle = D_{ij}\delta(t-t')
\end{equation}
and we can use the correlation functions of our noises, \eqref{Opticsnoise} and \eqref{Mechanicsnoise}, to get the matrix $D$ as 
\begin{equation}
D = 
\begin{pmatrix}
0 & 0 & 0 & 0 & 0 & 0 \\
0 & \gamma_1 (2\bar{n}_1 +1) & 0 & 0 & 0 & 0 \\
0 & 0 & 0 & 0 & 0 & 0 \\
0 & 0 & 0 & \gamma_2 (2 \bar{n}_2+1) & 0 & 0 \\
0 & 0 & 0 & 0 & 0 & \kappa/2 \\
0 & 0 & 0 & 0 & \kappa/2 & 0
\end{pmatrix}.
\end{equation}
Similarly, we can identify the covariance matrix $V$ in terms of our system coordinates $u(t)$ as 
\begin{equation}
V_{ij} = \frac{1}{2} \langle u_i(t) u_j(t') + u_j(t') u_i(t)\rangle.
\end{equation}
The steady-state expression for the covariance matrix can be found by solving the Lyapunov equation, 
\begin{equation}\label{Lyapunovequation}
AV + VA^T = -D
\end{equation}
under the condition that all eigenvalues (real parts) of the matrix $A$ are negative (equivalent to the the Routh-Hurwitz criterion for stability) \cite{Parks1981}. For our system this would result in a matrix of the form 
\begin{equation}
V = 
\begin{pmatrix}
\begin{matrix}
\langle \delta \hat{x}_1^2 \rangle & \langle \delta \hat{p}_1 \delta \hat{x}_1 \rangle \\
\langle \delta \hat{x}_1 \delta \hat{p}_1 \rangle & \langle \delta \hat{p}_1^2\rangle
\end{matrix} & \dots & \dots \\ 
\vdots & 
\begin{matrix}
\langle \delta \hat{x}_2^2 \rangle & \langle \delta \hat{p}_2 \delta \hat{x}_2 \rangle \\
\langle \delta \hat{x}_2 \delta \hat{p}_2 \rangle & \langle \delta \hat{p}_2^2\rangle
\end{matrix} & \dots \\
\vdots & \vdots & \ddots
\end{pmatrix}
\end{equation}
where the first four diagonal terms contain the autocorrelations of the position and momentum fluctuations of our two mechanical resonators. These are related to the effective thermal phonon number via 
\begin{equation}
\bar{n}_i^\mathrm{eff} = \frac{\langle \delta \hat{x}_i^2 \rangle + \langle \delta \hat{p}_i^2 \rangle - 1}{2}.
\end{equation}
Since our matrices $A$ and $D$ contain only known parameters, it is straightforward to numerically solve \eqref{Lyapunovequation} and obtain an expression for $\bar{n}_i^\mathrm{eff}$. In our notation, $\langle \delta \hat{x}_j^2 \rangle = \langle \delta \hat{p}_j^2 \rangle = 0.5$, which yields $\bar{n}_j^\mathrm{eff}=0$ for the mechanical ground state.

\subsection{Estimation of the optomechanical phase lag in other systems}

We can assess how the optomechanical phase lag would appear in different parameter regimes than the one we operate in, and specify some conditions necessary to see it. Firstly, in sideband-resolved systems, $\kappa \ll \omega_\mathrm{j}$, the optomechanically scattered light would be outside the cavity linewidth and would not form a standing wave, so the photons do not remain in the cavity for a significant time (cf.\ the system in \cite{Weaver2017} if their detuning would be small, $|\Delta|\ll \omega_j$). This would make the average time-delay between the transduced and local noises negligible, because it would be equal to the spatial separation of the resonators. In the opposite limit, $\kappa \gg \omega_\mathrm{m}$, the photons exit the cavity sufficiently quickly such that again the time delay becomes negligible. Ideally, both the laser carrier field $\omega_\mathrm{\ell}$ and the sidebands $\omega_\mathrm{\ell} \pm \omega_j$ fall within the cavity linewidth (i.e.\ $|\Delta| \ll \omega_j$) such that the effective coupling between the resonators is maximal. For example, if one operates with the laser carrier outside the cavity, the Stokes-rescattering of the anti-Stokes sideband (i.e.\ the second leg of process (4) in Fig.~\ref{FigS11}(a)) is weak.

Secondly, the optomechanical phase lag requires two mechanical resonators that are very close in frequency. Specifically, the mechanical (thermal) noise peaks must overlap to see interference between $\hat{\xi}_1$ and $\hat{\xi}_2$. This means that the difference in frequency must be smaller than the mechanical linewidths, $\omega_1 - \omega_2 < \gamma_1,~\gamma_2$, which is a challenging condition. It can be eased by operating in the regime of optomechanical cooling (as we do), where the linewidths are broadened.

Finally, the effective coupling rate must be significantly greater than the decoherence rate of the mechanical resonators. If the coupling rate is too small, the thermal decoherence of the resonators causes effectively a random phase relation between the local and transduced noise. On top of that, by utilizing a large cavity photon number, the $G_j$ become large and in that limit $M_1 (M_2) \rightarrow \frac{G_1}{G_2} (\frac{G_2}{G_1})$; the transduced noise becomes similar in size to the local noise and the interference between them is maximal.

The conditions on seeing interference due to optomechanical phase lag are three-fold:
\begin{itemize}
\item $\kappa > \omega_j$ and $|\Delta| \ll \omega_j$, we must be sideband-unresolved and operate the laser carrier within the cavity for the maximum coupling between the resonators.
\item $\omega_1 \simeq \omega_2$ (to within their effective linewidths), as otherwise the thermal noises do not overlap, and they cannot show interference.
\item The effective coupling rate ($\propto G_1G_2$) must be much greater than the effective decoherence rate ($\approx \gamma_1^\mathrm{eff},\gamma_2^\mathrm{eff}$), such that there is significant thermal noise that is transduced, and the phase relation between the resonator (noises) doesn't become random due to thermal decoherence.
\end{itemize}

We have summarized the relevant parameters for related work in Table~\ref{Tableestimation}, determined which of the above criteria they meet (\textcolor{green}\checkmark / \textcolor{red}{$\times$}) and how large the optomechanical phase lag $\phi_\mathrm{om}$ would be. Since $\omega_1 \simeq \omega_2$ is a challenging condition and subject to how the system is operated (i.e.\ mechanical frequencies can be changed by optomechanical cooling, thermal tuning, etc.), we have also calculated the phase lag for publications that are close to meeting this criterion (\textcolor{orange}{$\sim$} in Table~\ref{Tableestimation}). It is clear from the table that in all other publications to date, the optomechanical phase lag is small and therefore unlikely to be detected.\\

\begin{table}
\begin{center}
\begin{tabular}{llclcl}\hline
Ref. & $\omega_j/(2\pi)$ & $\omega_1\simeq\omega_2$? & $\kappa/(2\pi)$ & $\kappa > \omega_j$? & $\phi_\mathrm{om}$ ($\pi$ rad) \\
\hline
\cite{Xu2016} & \SI{788.04}{\kilo\hertz} & \textcolor{green}\checkmark & \SI{177}{\kilo\hertz} & \textcolor{red}{$\times$} & - \\
& \SI{788.49}{\kilo\hertz} & & & & \\
\hline
\cite{OckeloenKorppi2018} & \SI{10.0}{\mega\hertz} & \textcolor{red}{$\times$} & \SI{1.38}{\mega\hertz} & \textcolor{red}{$\times$} & - \\
& \SI{11.3}{\mega\hertz} & & & & \\
\hline
\cite{Lin2010} & \SI{13.6}{\mega\hertz} & \textcolor{green}\checkmark & \SI{6.9}{\giga\hertz} & \textcolor{green}\checkmark & $3.1\times 10^{-4}$ \\ 
\hline
\cite{Shkarin2014} & \SI{6.999}{\mega\hertz} & \textcolor{orange}{$\sim$} & $>$\SI{20}{\mega\hertz} & \textcolor{green}\checkmark & $0.055$ \\ 
& \SI{7.005}{\mega\hertz} & & & & \\
\hline
\cite{Zhang2012} & \SI{50.283}{\mega\hertz} & \textcolor{orange}{$\sim$} & $>$\SI{1}{\giga\hertz} & \textcolor{green}\checkmark & $2.7\times 10^{-3}$ \\ 
& \SI{50.219}{\mega\hertz} & & & & \\
\hline
\cite{Bagheri2013} & \SI{6.53}{\mega\hertz} & \textcolor{orange}{$\sim$} & $>$\SI{3}{\giga\hertz} & \textcolor{green}\checkmark & $3.0 \times 10^{-4}$ \\ 
& \SI{6.61}{\mega\hertz} & & & & \\
\hline
\cite{Sheng2020} & \SI{1.2}{\mega\hertz} & \textcolor{green}\checkmark & \SI{2}{\mega\hertz} & \textcolor{green}\checkmark & $0.095$ \\
\hline
\cite{Spethmann2016} & \SI{116.4}{\kilo\hertz} & \textcolor{orange}{$\sim$} & \SI{1.8}{\mega\hertz} & \textcolor{green}\checkmark & $0.01$ \\ 
& \SI{110.0}{\kilo\hertz} & & & & \\
\hline
This & \SI{149.90}{\kilo\hertz} & \textcolor{green}\checkmark & $250-600$~\si{\kilo\hertz} & \textcolor{green}\checkmark & $0.1 - 0.4$ \\
work & \SI{150.83}{\kilo\hertz} & & & & \\
\hline

\end{tabular}
\end{center}
\caption{Estimation of optomechanical phase lag in related works that meet (\textcolor{green}\checkmark) or come close to meeting (\textcolor{orange}{$\sim$}) the conditions to see optomechanical phase lag ($\omega_1 \simeq \omega_2$ and $\kappa > \omega_j$). Many related works operate in the resolved-sideband regime ($\kappa < \omega_j$) and have not been included.}
\label{Tableestimation}
\end{table}

\subsection{Exclusion of other interference mechanisms}

\subsubsection{Interference from effective mechanical frequency crossing}
\begin{figure*}
\includegraphics[width = 0.95\textwidth]{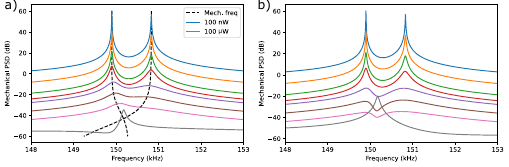}
\caption{Simulated PSD without (a) and with (b) optomechanical phase lag, for various powers (top lowest power, \SI{100}{\nano\watt}, bottom highest power, \SI{100}{\micro\watt}, offset vertically), with dashed black lines the effective mechanical frequencies for the different power curves.}
\label{FigureS6}
\end{figure*}

In certain optomechanical parameter regimes ($\kappa$, $\Delta$, $g_{0,j}$, input laser power $P_\ell$), the mechanical spectrum measured in our experiments can take a Fano-lineshape without the inclusion of optomechanical phase lag. As this looks qualitatively similar to the spectra we have reported in the main text, we will discuss how these phenomena are different. To illustrate this, we calculate the PSD using the parameters fitted to the data of Fig. 3(c) in the main text, shown in Table~\ref{Table1}. When the optical input power is increased, we can distinguish the different behavior with and without optomechanical phase lag in Fig.~\ref{FigureS6}.

For a low input laser power (\SI{100}{\nano\watt}, blue curves), the PSD looks the same without (a) and with (b) the optomechanical phase lag. The two mechanical resonances are clearly distinguishable as thermal peaks. As we increase the optical input power (orange through red curves), we see the peaks get broadened by optomechanical cooling. Without the optomechanical phase lag, Fig.~\ref{FigureS6}(a), the peaks continue to be broadened as we increase the power, until the effective mechanical frequencies cross (dashed lines) at higher powers. Around this crossing, the PSD changes and shows a dip left of a pronounced peak, forming a Fano lineshape. Comparing this with the PSD simulated with optomechanical phase lag, we see a dip appear in the curves of higher power ($\simeq$\SI{20}{\micro\watt}, purple through pink). This somewhat symmetric dip between the two mechanical peaks is notably absent when excluding optomechanical phase lag, and is a key feature of the spectra that we observe. For the highest power (grey curve), we enter the strong-coupling regime ($G_j > \kappa, \omega_j$) and we see a dark-mode peak appear while the bright mode becomes heavily damped.

The interference seen without optomechanical phase lag seems to be related to the effective mechanical frequencies crossing, and can be compared to the work of Ref.~\cite{Lin2010}. Though the mechanism behind the interference effect is different in that work, their mechanical modes cross in a similar manner to the simulations shown in Fig.~\ref{FigureS6}. As can be seen from the data reported in Fig.~4(a) of the main text (where we increase the laser input power while keeping $g_{0,1}/g_{0,2}$ constant), mechanical frequencies have not crossed and the curves resemble those of Fig.~\ref{FigureS6}(b) much more than they resemble the curves of Fig.~\ref{FigureS6}(a). Furthermore, the optomechanical parameters $\kappa$, $\Delta$, $g_{0,j}$ and $\omega_j$ follow from independent measurement of the cavity and OMIT (see below). Based on these considerations, the interference effect we see is significantly different from the one that originates from the effective mechanical mode crossing.

\subsubsection{Interference from direct mechanics-mechanics coupling}
\begin{figure}
\includegraphics[width = 1.0\textwidth]{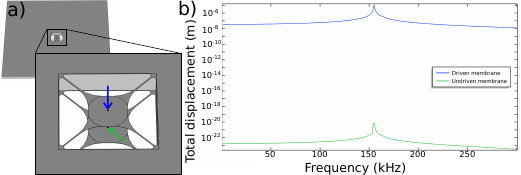}
\caption{a) COMSOL\textsuperscript{\tiny\textregistered} model of our double-membrane device. The inset shows an area around the membranes and the hole through the chip, with the blue arrow denoting the location of a simulated driving force and the green arrow denoting the readout point in the center of the other membrane. b) Frequency-response simulation showing negligible displacement of the undriven membrane.}
\label{FigureS5}
\end{figure}

If there is a direct mechanical coupling between the resonators, an interference effect can also be seen in the mechanical PSD~\cite{Lin2010}. Such a coupling would be described by a term of the form $c \hat{x}_1 \hat{x}_2$ in the Hamiltonian ($c$ some coupling constant), which is absent in our \eqref{Hamiltonian}. To validate this, we estimate the direct mechanical coupling between the membranes (through the Si chip) via a finite element method (FEM) simulation in COMSOL\textsuperscript{\tiny\textregistered}. Our model consists of two 2D-shell physics nodes representing the suspended membrane and the SiN layer on each side of the chip, while a 3D solid mechanics physics node represents the bulk of the Si chip (Fig.~\ref{FigureS5}(a)). The in-plane stress in the SiN layer is included in the model, and we capture the stress redistribution from the release step of the fabrication by including a stationary step in the model to ensure the stress distribution and geometry of our model match the physical sample.  We ensure the mesh in the two membranes is identical, and sufficiently fine to have a converging solution.

We apply a \SI{1}{\micro\newton} harmonic perturbation to the middle of one of the membranes (Fig.~\ref{FigureS5}(a), blue arrow), and obtain the displacement both at this center point and the center point of the other, undriven membrane (green arrow). The difference in displacement amplitude gives an indication of the coupling strength between the membranes. From the 15 orders-of-magnitude difference in displacement plotted in Fig.~\ref{FigureS5}(b), we can confidently say that the direct mechanical coupling is negligibly small. From that, we conclude that the interference mechanism studied in~\cite{Lin2010} is not present in our system.

\subsubsection{Interference from back-action cancellation}
There are several publications describing interference in optomechanical systems through back-action cancellation, e.g.~\cite{Caniard2007,Dobrindt2010,Yanay2016}. In general, this requires either multiple optical cavities/optical modes, or a second beam to drive the system. In this work, we explicitly only have a single optical cavity mode, which we verify by scanning the laser wavelength across the cavity resonance; we see a single resonance within the scanning window ($\pm$\SI{60}{\mega\hertz}). There are other modes from the Fabry-Pérot cavity, the longitudinal modes are separated by the free spectral range (FSR, \SI{3}{\giga\hertz}) and the transversal modes are separated by $>$\SI{100}{\mega\hertz}. Both of these are considerably larger than the mechanical frequency ($\sim$\SI{150}{\kilo\hertz}), which precludes them from being relevant in back-action cancellation schemes such as in Ref.~\cite{Dobrindt2010}.

The mode-matching is about 92\% to the longitudinal mode, so the transversal modes are considerably smaller (measured in the empty cavity). The largest contribution is the second transversal mode, due to a slight mismatch in beam size between the incident laser beam and the cavity. This further allows us to exclude other optical modes as a mechanism for interference.

The incident laser beam consists of a single tone around \SI{1550}{\nano\meter}, with two sidebands at \SI{30}{\mega\hertz} generated for the Pound-Drever-Hall scheme. During operation, these sidebands are well outside the cavity window, and do not match with other frequencies of the system (i.e. $\omega_j$). Thus we can also exclude back-action cancellation schemes relying on optical drives such as \cite{Caniard2007}. For the OMIT measurements (next section), we do use a drive tone that we sweep across the mechanical frequency, but this is turned off when measuring the mechanical spectra that show optomechanical phase lag and the resulting interference.\\

\subsection{Multi-mode OMIT}\label{OMITsection}
We use optomechanically induced transparency (OMIT) \cite{Weis2010} as a way to fit and extract the optomechanical parameters of our system. Since we have multiple mechanical resonators and cannot operate our cavity in the sideband-resolved limit, we adopt the approach of Ref. \cite{Nielsen2017} to fit our measured OMIT curves. We start by neglecting all input noises, and keeping only the strong pump term $\sqrt{\kappa_\mathrm{e}} a_\mathrm{d}$. We choose a phase reference such that the average cavity field $\langle \hat{a}\rangle$ is real and positive, such that $G_1$ and $G_2$ are also real and positive. We use the equations
\begin{equation}
\begin{aligned}
\langle \hat{x}_j \rangle &= \chi_j(\omega) G_j \left( \langle \hat{a} \rangle + \langle \hat{a} \rangle^* \right) \\
\langle \hat{a} \rangle &= \chi_\mathrm{c}(\omega) \left( \sum_{j=1,2} i G_j \langle \hat{x}_j \rangle  + \sqrt{\kappa_\mathrm{e}} a_\mathrm{d}\right) \\
\langle \hat{a} \rangle^* &= \chi_\mathrm{c}^* (-\omega) \left( \sum_{j = 1,2} -i G_j \langle \hat{x}_j \rangle + \sqrt{\kappa_\mathrm{e}} a_\mathrm{d}^*\right).
\end{aligned}
\end{equation}
We solve this set of equations for the cavity field, and at the output we measure, $S(\omega) = \sqrt{\kappa_\mathrm{e}} (\langle \hat{a} \rangle + \langle \hat{a} \rangle^*)$, and obtain
\begin{equation}\label{OMITspectrum}
S(\omega) = \frac{\kappa_\mathrm{e} \left( \chi_\mathrm{c}(\omega) a_\mathrm{d} + \chi_\mathrm{c}^*(-\omega) a_\mathrm{d}^* \right)}{1 - i\left(\chi_\mathrm{c}(\omega) -\chi_\mathrm{c}^*(-\omega)\right) \left( \chi_1(\omega) G_1^2 + \chi_2(\omega) G_2^2 \right)}.
\end{equation}
This result is similar to the one obtained in Ref.~\cite{Nielsen2017}. It contains terms both at $+\omega$ and $-\omega$, which represent the anti-Stokes and Stokes signal generated by our mechanical resonators, which we need to take into account due to the level of sideband resolution in our system ($\kappa \gtrsim \omega_j$). When comparing to the expressions in the fully sideband-resolved limit \cite{Weis2010}, we see that those do not contain both of these terms.

\begin{figure}
\centering
\includegraphics[width = 0.65\textwidth]{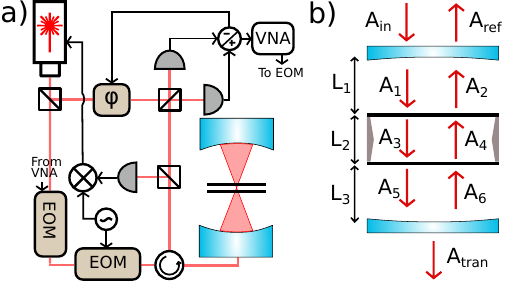}
\caption{a) Setup to measure OMIT from our system. A VNA drives an additional phase EOM in the frequency range close to our mechanical resonances, and reads the reflected homodyne signal. b) Cavity fields for our Fabry-Pérot cavity containing the double-membrane system.}
\label{SetupOMIT}
\end{figure}

To detect our OMIT signal, we add an additional electro-optic modulator (EOM) to our setup, and we connect a vector network analyzer (VNA) to it to provide a frequency sweep of the drive together with the read-out. We connect the VNA in place of the spectrum analyzer (SA) to the home-built homodyne detector, as shown schematically in Fig.~\ref{SetupOMIT}. To measure OMIT, we lock the frequency of our laser to our cavity, and we turn on the small drive from the VNA. Due to the small linewidth of our mechanical resonators ($\simeq$1~Hz without optomechanical cooling), the filter bandwidth of the VNA is set very narrowly and we integrate for several minutes.

In OMIT measurements shown in the main text, we see an additional feature not expected from the spectrum derived in \eqref{OMITspectrum}. This happens at exactly the mechanical frequencies of the resonators, and takes the shape of a peak in the OMIT curve. Because of the frequency at which this happens, we ascribe this to mechanical (thermal) noise from the resonators that we neglected in the derivation above. For a sufficiently strong pump $\hat{a}_\mathrm{d}$, the mechanical noise would be negligible, but such a pump would also affect our system. The frequency locking of the cavity is experimentally challenging, and a pump strong enough to neglect the mechanical noise would not allow us to obtain a stable lock. For the fits of the OMIT spectrum of \eqref{OMITspectrum}, the frequency of these mechanical noise features make it difficult to exclude them, leading to some uncertainty in the fit parameters, in particular the detuning $\Delta$.

\begin{figure}
\centering
\includegraphics[width = 0.7\textwidth]{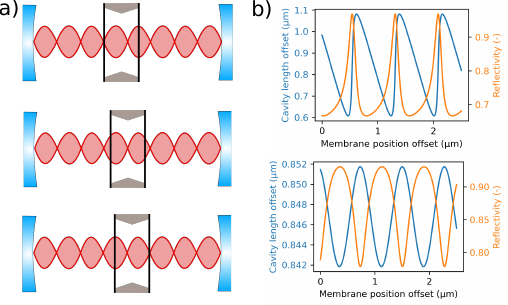}
\caption{a) Sketch of control of (linear) optomechanical coupling $g_0$ by changing the position of the membranes within the cavity. From top to bottom:\ left resonator maximally coupled, right resonator not (linearly) coupled; both resonators approximately equally coupled; and left resonator not coupled, right resonator fully coupled. b) Simulated dispersion curve of cavity resonance versus chip position, and the associated reflectivity of the (lossy) cavity. Top panel for the slightly off-resonant case in this work, bottom panel for the case where the intra-membrane cavity is at resonance.}
\label{Schematiccavityfield}
\end{figure}

\subsection{Fabry-Pérot cavity with two lossy reflecting membranes}\label{Couplingvariation}
To model the behavior of our system, we use a known model~\cite{Li2016} to describe our cavity in terms of the optical field amplitudes. The reason for this is twofold:\ Firstly, we control the optomechanical coupling of both of the membranes by controlling the position of the membrane-chip and the length of the cavity. Together with the (tunable) wavelength of the light that we send in, this gives us control over the resonance conditions of the three sub-cavities in our system. Secondly, by including lossy membranes in the model, we can take into account how the cavity linewidth changes as a function of membrane position. This gives us bounds on the optomechanical phase lag that we can expect.

We use the model
\begin{equation}\label{Cavityfields}
\begin{aligned}
A_1 &= i t A_\mathrm{in} + r A_2 e^{ikL_1} \\
A_2 &= i t_m A_4 e^{ikL_2} - r_m A_1 e^{ikL_1} \\
A_3 &= i t_m A_1 e^{ikL_1} - r_m A_4 e^{ikL_2} \\
A_4 &= i t_m A_6 e^{ikL_3} - r_m A_3 e^{ikL_2} \\
A_5 &= i t_m A_3 e^{ikL_2} - r_m A_6 e^{ikL_3} \\
A_6 &= r A_5 e^{ikL_3} \\
A_\mathrm{ref} &= i t A_2 e^{ikL_2} + r A_\mathrm{in} \\
A_\mathrm{tran} &= i t A_5 e^{ikL_3}
\end{aligned}
\end{equation}
with the field amplitudes $A_\mathrm{in}$, $A_\mathrm{refl}$, $A_\mathrm{tran}$ and $A_{1-6}$ and lengths $L_1$, $L_2$ and $L_3$ defined as in Fig.~\ref{SetupOMIT}(b). Furthermore, we have mirror reflectivity $r^2 = 99.995\%$ and transmissivity $t^2 = 1- r^2$ and membrane reflectivity $r_m^2 \approx 35\%$ and transmissivity $t_m^2 \approx 65\%$. This means we have lossless mirrors, which is a relatively good approximation for our setup. For the membranes, the loss due to absorption is much smaller than the losses due to e.g.\ scattering~\cite{Gaertner2018}, imperfect alignment, fabrication imperfections, etc., so it can be ignored. We can take non-absorption losses into account by reducing the reflectivity and transmissivity slightly from their stated values, using a value of 0.98 ($= r_m^2 + t_m^2$) for the losses. This reduces the transmission of the cavity, but does not change the shape of the dispersion curve, i.e. it does not affect the optomechanical coupling rate. We have a photonic crystal pattern on our membranes~\cite{Gaertner2018} that changes the index of refraction away from that of bare SiN, which differs from the assumptions in Ref.~\cite{Li2016}.  

\begin{figure*}
\includegraphics[width = 1.0\textwidth]{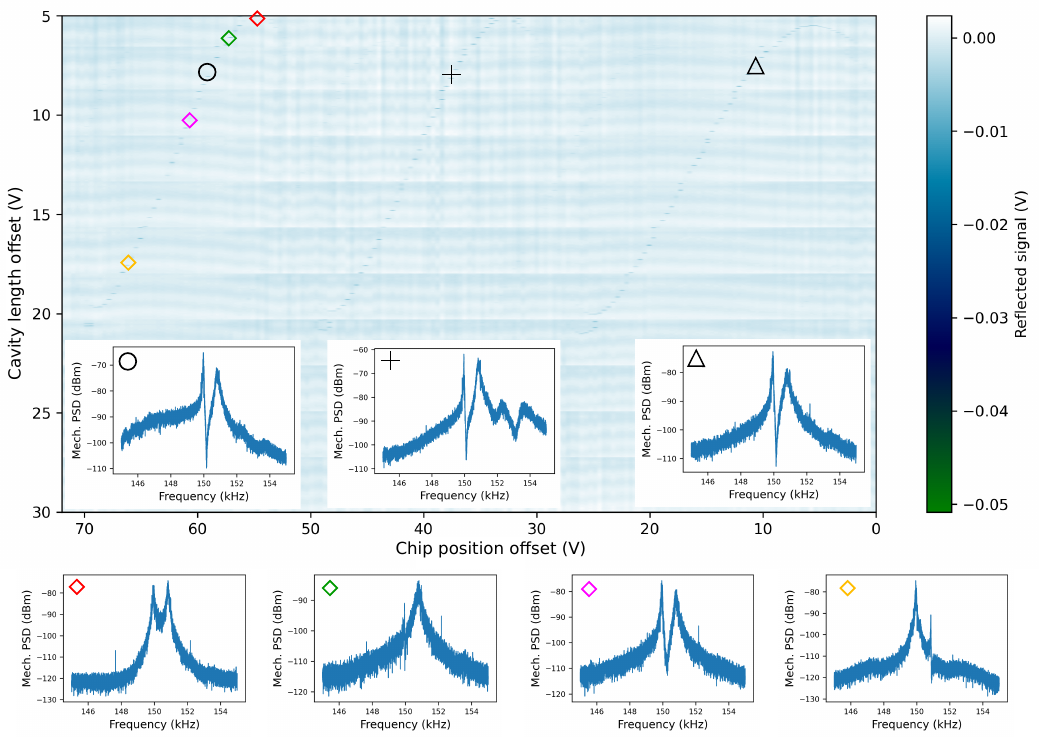}
\caption{Cavity resonance as a function of membrane position. The main figure shows the dispersion curve of the resonance, which shifts in frequency ($\omega_\mathrm{c} \propto V_\mathrm{mirror}$) when the chip position is varied. Insets with black symbols show the homodyne spectrum at similar points on different branches of the dispersion curve. Insets with colored diamonds show the homodyne spectrum at different points on the same branch of the dispersion curve. Horizontal stripes are artifacts of the measurement.}
\label{Dispersioncurve}
\end{figure*}

We analytically solve \eqref{Cavityfields} for the fields $A_{1-6}$, $A_\mathrm{in}$, $A_\mathrm{refl}$ and $A_\mathrm{tran}$ and perform numerical calculations using the resulting equations. We vary the position of the membrane chip and calculate the position of the cavity mirrors required to see a cavity resonance for every position of the membrane. This gives a dispersion curve of the cavity resonance, Fig.~\ref{Schematiccavityfield}(b) (blue lines), in exactly the same way as we perform experimentally. This dispersion curve takes a sinusoidal shape when the inter-membrane cavity is resonant (bottom plot), but is skewed when we are off-resonant with the inter-membrane cavity (top plot). Since the membrane reflectivity is much lower than the mirror reflectivity, the cavity can still be resonant even when the inter-membrane cavity is off-resonance. When comparing these numerical simulations to the measurements in Fig.~\ref{Dispersioncurve}, we see from the skewness of the dispersion curve that we are not on resonance with the inter-membrane cavity. The regime of inter-membrane cavity resonance is of significant interest due to enhancement of the optomechanical coupling~\cite{Li2016,Piergentili2018,Newsom2020}, but unfortunately out of reach due to limited laser tunability.

In Fig.~\ref{Schematiccavityfield}(b), we also plot the reflectivity of the cavity as a whole (orange lines). The plotted reflectivity gives the depth of the dip in cavity reflection at the resonance, i.e. the minimum reflection. The closer this value is to 1, the less we see of the cavity resonance due to losses in the cavity. We utilize this as a simple model to estimate the expected cavity linewidth to calculate the optomechanical phase lag. We estimate a lower (\SI{275}{\kilo\hertz}) and upper (\SI{600}{\kilo\hertz}) bound for the linewidth at the point of the lowest reflectivity, based on averaged measurements of the cavity linewidth over various measurement runs. Then, we scale these linewidths based on the reflectivity calculated and plotted in Fig.~\ref{Schematiccavityfield}(b) to incorporate the effect of cavity losses on the expected optomechanical phase lag, thereby neglecting any contribution from the cavity lock.

When varying the position of the chip, we can choose for either of the membranes to be at a node/antinode of the field, as depicted in Fig.~\ref{Schematiccavityfield}(b). As we sweep the position of the membrane, our resonance traces out the dispersion curve and we can smoothly vary the coupling to either membrane. We do so in a measurement, plotted in the main window of Fig.~\ref{Dispersioncurve}. At several points (colored diamonds) we measure the PSD and we can deduce the coupling situation of each of the membranes from that. For the green (orange) diamond, we are predominantly coupled to the higher (lower) frequency resonator, given by its broadened linewidth in the noise spectrum. In the purple diamond case, we are coupled approximately equally to both membranes. As is visible in the main window, these three colored diamonds are in the mostly-linear regime where the cavity frequency shift is linear with membrane position, hence our optomechanical coupling is linear. The red diamond shows a measurement done in the quadratic regime, which shows a markedly different noise spectrum (no interference). 

In the main window of Fig.~\ref{Dispersioncurve}, we can see approximately three periods of the dispersion curve. When comparing the noise spectra at identical points on the different curves (black circle, cross and triangle), we see a difference in features around 147~kHz and 153~kHz, which we ascribe to some other mechanical mode (not necessarily of our membranes). The difference in these features between the three periods of the dispersion curve we attribute to different amplitudes in the solutions to the field equations, \eqref{Cavityfields}. All measurements done in the main text were performed at a branch of the dispersion curve where both noise features (147 and 153~kHz) were mostly absent.

\subsection{Analytical treatment of cooperativity competition on mechanical dissipation}\label{CCderivation}
To clarify the effect of the cooperativity competition, we derive an analytical expression using some simplifying assumptions. In the derivation below, we have chosen a phase reference such that $\langle \hat{a} \rangle$ (and thus $G_j$) is real and positive. In all the numerical evaluations throughout this work, we have kept the complex behaviour of $G_j$.
From the real part of $\left(\chi_1^{\rm eff}\right)^{-1}$ in \eqref{chi1effs}, we can extract the effective mechanical frequency, where we can recognize the so-called "optical-spring" effect, given by
\begin{equation}\label{w1eff}
\omega_{1}^{\rm eff} (\omega) = \Bigg[  \omega_1^2 +   \frac{2 G_1^2 \Delta  \omega_1 \, \big[ \AAA (\omega) - (\Delta^2 +\kappa^2/4 - \omega^2)  \big]}{ \big[ \AAA (\omega) {-} (\Delta^2 {+} \kappa^2/4 {-} \omega^2) \big]^2 + \big[ \kappa \omega {+} \BB(\omega) \big]^2 }  \Bigg]^{\frac{1}{2}}, 
\end{equation}
where 
\begin{equation}
\begin{aligned}
\AAA (\omega) &:=   \frac{2 G_2^2 \Delta \omega_2 (\omega_2^2 - \omega^2) }{ (\omega_2^2 - \omega^2)^2 + \gamma_2^2 \omega^2 }   , \\
\BB (\omega) &:=   \frac{2 G_2^2 \Delta \omega_2  \gamma_2 \omega }{ (\omega_2^2 - \omega^2)^2 + \gamma_2^2 \omega^2 } .  
\end{aligned}
\end{equation}
From the imaginary part of $\left(\chi_1^{\rm eff}\right)^{-1}$ in \eqref{chi1effs}, we can extract the effective mechanical damping rate
\begin{equation}\label{gama1}
\gamma_{1}^{\rm eff} (\omega) = \gamma_{1} +  \frac{\omega_1}{\omega}  \frac{2 G_1^2 \Delta \, \big[ \kappa \omega +\BB(\omega) \big]}{ \big[ \AAA (\omega) {-} (\Delta^2 {+} \kappa^2/4 {-} \omega^2) \big]^2 + \big[ \kappa \omega {+} \BB(\omega) \big]^2 }. 
\end{equation}
\eqref{w1eff} and \eqref{gama1} clearly reveal that both the effective frequency and damping rate of the first mechanical resonator are modified by the presence of the second mechanical resonator, reflected in the fact that $\AAA (\omega), \BB (\omega) \ne 0$ when $G_2 \ne 0$. Instead, if we take $G_2 = 0$, \eqref{w1eff} and \eqref{gama1} become exactly the same form as those reported in Ref.~\cite{Genes2008} for a single resonator.

We now focus on the effect of the coupling to the second mechanical resonator on the damping rate of the first mechanical resonator. The expression of $\gamma_{1}^{\rm eff} (\omega)$ is still quite involved, but it takes a simpler form under specific interesting conditions. By assuming the mechanical resonators with equal frequencies $\omega_1 = \omega_2 \equiv \omega_0 $, working in the optimal mechanical cooling regime $\Delta = \omega_0$ (the cavity is resonant with two anti-Stokes sidebands), and looking at $\omega= \omega_0$ in the spectrum, we obtain
\begin{equation}
\AAA (\omega_0)  =0,  \,\,\,\,\,\, \BB (\omega_0) = \frac{2 G_2^2 \omega_0}{\gamma_2},
\end{equation}
and thus
\begin{equation}\label{gamma1G2}
\gamma_{1}^{\rm eff} (\omega_0) = \gamma_{1} +  \frac{2 G_1^2 \Big( \kappa +\frac{2 G_2^2}{\gamma_2} \Big)} {\frac{\kappa^4}{16 \omega_0^2} + \Big( \kappa +\frac{2 G_2^2}{\gamma_2} \Big)^2  }. 
\end{equation}
By assuming a large cooperativity of the second mechanical resonator, $C_2 = 2G_2^2/(\kappa \gamma_2) \gg 1$, we achieve
\begin{equation}
\gamma_{1}^{\rm eff} (\omega_0) \simeq  \gamma_{1} +  \frac{2 G_1^2 \kappa  C_2 } {\kappa^2  \Big( \frac{\kappa^2}{16 \omega_0^2} +C_2^2 \Big)} ,
\end{equation}
and by further assuming $\kappa < 4 \omega_0$, thus $(\kappa /4 \omega_0)^2 < 1 \ll C_2^2$, we obtain a rather simple expression
\begin{equation}\label{gamma1}
\gamma_{1}^{\rm eff} (\omega_0) \simeq  \gamma_{1} +  \frac{ 2 G_1^2 } {\kappa C_2} =  \gamma_{1} \Bigg( 1+ \frac{C_1}{C_2} \Bigg) ,
\end{equation}
where $C_1 = 2 G_1^2/(\kappa \gamma_1)$ is the cooperativity of the first mechanical resonator, which does not have to be large to derive the above equation. It is interesting to compare $\gamma_{1}^{\rm eff}$ with and without the second mechanical resonator.  By taking $G_2=0$ in \eqref{gamma1G2} (only one mechanical resonator is coupled to the cavity), we have 
\begin{equation}\label{gamma1G1}
\gamma_{1}^{\rm eff} (\omega_0) = \gamma_{1} +  \frac{2 G_1^2 } {\kappa \Big( \frac{\kappa^2}{16 \omega_0^2} + 1 \Big)  },
\end{equation}
and it becomes the well-known result in the resolved sideband limit $\kappa \ll  \omega_0$~\cite{WilsonRae2007,Marquardt2007,Genes2008},
\begin{equation}\label{gamma1G111}
\gamma_{1}^{\rm eff} (\omega_0) \simeq \gamma_{1} (1 + C_1).
\end{equation}
Note that here the condition $\kappa \ll  \omega_0$ is more demanding on the cavity linewidth than $\kappa < 4 \omega_0$ used for deriving \eqref{gamma1}, because the latter is only used to keep $(\kappa /4 \omega_0)^2 < 1 \ll C_2^2$. Comparing the damping rates with and without the second mechanical resonator, i.e. \eqref{gamma1} and~\eqref{gamma1G111}, we see that the effective damping rate $\gamma_{1}^{\rm eff}$ is significantly reduced due to the presence of the second mechanical resonator because a large $C_2 \gg 1$ is assumed in \eqref{gamma1}, and for the special case $C_1=C_2$, it reduces to twice its natural damping rate $\gamma_{1}^{\rm eff} = 2 \gamma_1$. 

Similarly, owing to the symmetry of the two mechanical resonators, we obtain the effective damping rate of the second mechanical resonator
\begin{equation}\label{gamma2}
\gamma_{2}^{\rm eff} (\omega_0) \simeq  \gamma_{2} \Bigg( 1+ \frac{C_2}{C_1} \Bigg) ,
\end{equation}
under the condition $C_1 \gg 1$. \eqref{gamma1} and \eqref{gamma2} are the main results of the work. They reveal a dissipation competition mechanism between the two mechanical resonators:\ the effective damping rate of each mechanical resonator is reduced by the presence of the other mechanical resonator, with the extent depending on the ratio of their cooperativities $C_1/C_2$, and the mechanical resonator with a larger cooperativity dissipates faster, i.e.\
\begin{equation}
C_1 > C_2 \, \Rightarrow \, \gamma_{1}^{\rm eff} (\omega_0) > \gamma_{2}^{\rm eff} (\omega_0),
\end{equation}
if the two natural damping rates are assumed equal $ \gamma_{1}= \gamma_{2}$ (in our system, these two damping rates are very close). This can be understood intuitively:\ both the mechanical resonators dissipate through the same optical channel and the one that is more strongly coupled to the optical field takes advantage in dissipating energy via light into the environment. We call this phenomenon cooperativity competition on the mechanical dissipation.

\subsection{Mechanical noise cancellation for sensors}
The ability to cancel mechanical (thermal) noise by insertion of a second mechanical resonator hints at potential applications in sensing. The straightforward way of implementing this would be to consider a resonant signal (force) $\hat{q}_1$ acting on the position fluctuations of one of the membranes. An example of this would be a laser beam at an incident angle such that it does not form a cavity mode, which is modulated at a specific (signal) frequency. This can be described by modifying \eqref{p_j}-\eqref{x_j} to

\begin{equation}
\begin{aligned}
\delta \hat{p}_j &= -j \frac{\omega}{\omega_j} \delta \hat{x}_j, \\
\delta \hat{x}_j &= \chi_j(\omega) \left[ G_j^* \delta \hat{a} + G_j \delta \hat{a}^\dagger  + \hat{\xi}_j + \hat{q}_1\right]. \\
\end{aligned}
\end{equation} 
One can retrieve the solutions from earlier in this document by substituting $\hat{\tilde{\xi}}_1 = \hat{\xi}_1 + \hat{q}_1$, which shows that the signal we would want to detect is canceled by the mechanical noise cancellation, similar to the thermal noise. That is, in the transparency window (where the mechanical noise is canceled), the signal would appear above the thermal noise with the same prominence as outside the transparency window. 

\begin{figure}
\includegraphics[width = 0.5\textwidth]{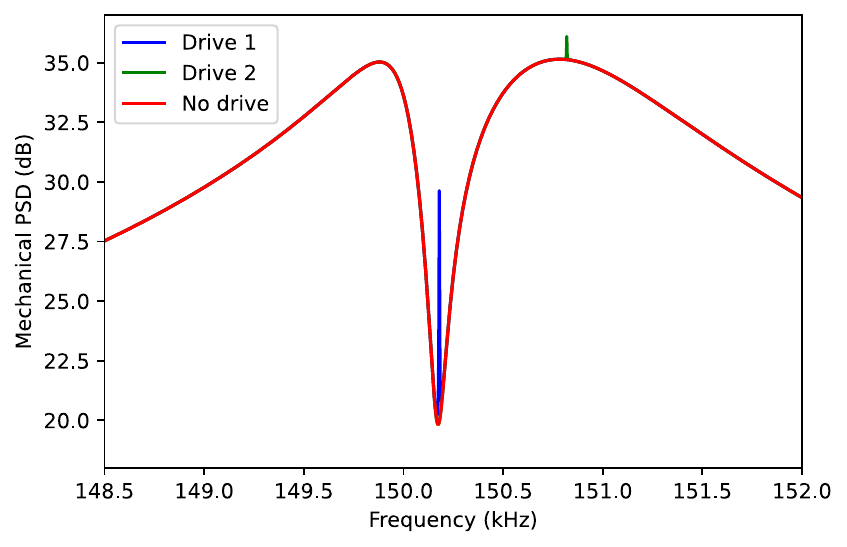}
\caption{PSD with optical signal in cancellation window (blue), outside cancellation window (green), and without driving (red). The contrast between the signal peak and (thermal) noise floor is enhanced when it falls within the cancellation window.}
\label{FigureSensitivity}
\end{figure}

Conversely, if we work with a signal that is present in the optical mode rather than on the mechanics side, we introduce the $\hat{q}_1$ signal in \eqref{aa}-\eqref{ad} instead,
\begin{equation}
\begin{aligned}
\delta \hat{a} &= \chi_c (\omega) \left( \sum_{j=1,2} i G_j \delta \hat{x}_j + \sqrt{\kappa}\hat{a}^\mathrm{in} + \hat{q}_1 \right)\\
\delta \hat{a}^\dagger &= \chi_c^* (-\omega) \left( \sum_{j=1,2} -i G_j^* \delta \hat{x}_j + \sqrt{\kappa} \hat{a}^{\mathrm{in},\dagger} + \hat{q}_1^\dagger \right).
\end{aligned}
\end{equation}
We can take it through the same process as described earlier in this work to calculate the resulting PSD. We assume a narrow-frequency signal with a Lorentzian distribution centered at frequency $\omega_q$, with linewidth $\gamma_q$ and with power $P_q$. If the signal is in the transparency window (Fig.~\ref{FigureSensitivity} blue curve), it is much better resolved above the noise than if the signal is present outside the transparency window (green curve). We have used realistic parameters close to those of Table~\ref{Table1} to obtain these spectra. A particular use-case for such a signal could be in a gravitational wave interferometer \cite{LIGO2015} if one wants to detect a signal that is at the same frequency as a mechanical mode of that system. In that case, the fluctuations originating from (unwanted) mechanical modes can be suppressed in a specific frequency range by the inclusion of this second resonator in the cavity.

\end{document}